\documentclass[reqno]{amsart}
\usepackage{amsfonts,amsmath,amssymb,bbm}
\usepackage{url}
\usepackage{bbm}
\usepackage{verbatim}

\newcommand{\etalchar}[1]{$^{#1}$}
\providecommand{\bysame}{\leavevmode\hbox
to3em{\hrulefill}\thinspace}
\providecommand{\MR}{\relax\ifhmode\unskip\space\fi MR }
 \providecommand{\href}[2]{#2}

\newcommand{\eq}{\begin{equation}}
\newcommand{\en}{\end{equation}}
\newcommand{\eqa}{\begin{eqnarray}}
\newcommand{\ena}{\end{eqnarray}}
\newcommand{\eqas}{\begin{eqnarray*}}
\newcommand{\enas}{\end{eqnarray*}}
\newcommand{\ra}{ {\rightarrow} }
\newcommand{\Ra}{ {\Rightarrow} }
\def\bes{\begin{subarray}{c} }
\def\es{\end{subarray} }

\newcommand{\npp}{\textsc{Npp}}

\newcommand{\abs}[1]{\ensuremath{\left\vert #1 \right\vert}}
\providecommand{\binom}[2]{\left(#1\atop#2\right)} \makeatletter
\def\erf{\mathop{\operator@font erf}\nolimits}
\def\erfc{\mathop{\operator@font erfc}\nolimits}
\def\argmax{\mathop{\operator@font argmax}\nolimits}
\def\sinc{\mathop{\operator@font sinc}\nolimits}
\def\rect{\mathop{\operator@font rect}\nolimits}
\def\poly{\mathop{\operator@font poly}\nolimits}
\@ifundefined{widetext}{}{}
\makeatother

\def\sumbeta{\!\!\mathop{\,\,{\sum}^{(\beta)}}}

\def\mb#1{\mathbf{#1}}
\def\bss{\boldsymbol{\sigma}}
\def\bsd{{\boldsymbol{\delta}}}
\def\R{{\mathbb R}}
\def\I{{\mathbb I}}
\def\PR{{\mathbb P}}
\def\EX{{\mathbb E}}

\newtheorem{theorem}{Theorem}[section]
\newtheorem{lemma}[theorem]{Lemma}

\newtheorem{conjecture}[theorem]{Conjecture}

\theoremstyle{definition}
\newtheorem{definition}[theorem]{Definition}

\theoremstyle{remark}
\newtheorem{remark}[theorem]{Remark}

\def\mb#1{\mathbf{#1}}

\numberwithin{equation}{section}

\newcommand{\rhat}{\hat{\rho}}

\begin{document}

\title[Proof of the local REM conjecture for number partitioning
I]{Proof of the local REM conjecture for number partitioning
I: Constant energy scales}

\author{Christian Borgs$^1$, Jennifer Chayes$^1$, Stephan Mertens$^2$,
        Chandra Nair$^3$}

\address{$^1$Microsoft Research, One Microsoft Way, Redmond, WA 98052}

\address{$^2$Inst.\ f.\ Theor.\ Physik,
    Otto-von-Guericke Universit\"at, PF~4120, 39016 Magdeburg, Germany}

\address{$^3$Dept. Of Elec. Eng., Stanford University, CA 94305}


\date{April 21, 2005}

\begin{abstract}
The number partitioning problem is a classic problem of
combinatorial optimization in which a set of $n$ numbers
is partitioned into two subsets such that the sum of
the numbers in one subset is as close as possible to the
sum of the numbers in the other set. When the $n$ numbers
are i.i.d.~variables drawn from some distribution,
the partitioning problem turns out to be equivalent to a
mean-field antiferromagnetic Ising spin glass.  In the spin
glass representation, it is natural to define energies
-- corresponding to the costs of the partitions, and
overlaps -- corresponding to the correlations between
partitions.  Although the energy
levels of this model are {\em a priori} highly correlated, a
surprising recent conjecture asserts that the energy spectrum of
number partitioning is locally that of a random energy model (REM):
the spacings between nearby energy levels are uncorrelated.  In
other words, the properly scaled energies converge to a Poisson
process.   The conjecture also asserts that the corresponding spin
configurations are uncorrelated, indicating vanishing overlaps in
the spin glass representation.  In this paper, we prove these two
claims, collectively known as the local REM conjecture.
\end{abstract}

\maketitle

\section{Introduction}
\label{sec:intro}

The study of typical
properties of random instances of combinatorial problems
has recently been the focus of much interest in the theoretical
computer science, discrete mathematics and statistical physics
communities.  Many
of these problems turn
out to be closely related to disordered problems in statistical
physics \cite{tcs-phasetransitions,mezard:03} -- a
connection which has motivated a host of interesting conjectures.
In this paper, we establish one of these conjectures:
the local REM property of the random number partitioning
problem (\npp).

The non-random \npp\ is one of the classic NP-complete problems
of combinatorial optimization,
closely related to other classic problems such as
bin packing, multiprocessor scheduling, quadratic
programming and knapsack problems \cite{garey:johnson:79, ausiello:etal:99}.
In
addition to its theoretical significance, the \npp\  has many
applications including task scheduling and the minimization of VLSI
circuit size and delay \cite{coffman:lueker:91,tsai:92}, public key
cryptography \cite{merkle:hellman:78,odlyzko:91}, and, more
amusingly, choosing teams in children's baseball games
\cite{hayes:npp}.

A fixed instance of the \npp\ is defined as follows:
Given $n$ numbers $X_1,X_2,$ $\ldots, X_n$, we seek a partition of
these numbers into two subsets such that the sum of numbers in one
subset is as close as possible to the sum of numbers in the other
subset.  Each of the $2^n$ partitions can be encoded as
 $\bss \in  \{-1,+1\}^n$, where $\sigma_i=1$ if
$X_i$ is put in one subset and $\sigma_i=-1$ if $X_i$ is put in the
other subset; in the
physics literature, such partitions $\bss$ are identified with {\em Ising
spin configurations}.  The cost function to be minimized over all spin
configurations $\bss$ is therefore the {\em energy}
\begin{equation}
  \label{ene}
  E(\bss) = \frac{1}{\sqrt{n}}\left|\sum_{s=1}^n \sigma_s X_s \right|,
\end{equation}
where we have inserted a factor $1/\sqrt{n}$ to simplify the
equations in the rest of the paper.

Note that the spin configurations $\bss$ and $-\bss$ correspond to
the same partition and therefore of course have the same energy.
Thus there are $N = 2^{n-1}$ distinct partitions and at most $N$
distinct energies. The lowest of these $N$ energies is the {\em
ground state energy} of the model. The {\em energy spectrum} is the
sorted increasing sequence $E_1,...,E_N$ of the energy values
corresponding to these $N$ distinct partitions.  Let
$\bss^{(1)},\dots,\bss^{(N)}$ be configurations corresponding to
these ordered energies. The {\em overlap} between the configurations
$\bss^{(i)}$ and $\bss^{(j)}$ is defined as
\begin{equation}
\label{overlap} q(\bss^{(i)},\bss^{(j)})= \frac 1n
\sum_{s=1}^n\sigma^{(i)}_s\sigma^{(j)}_s .
\end{equation}

One often studies random instances of the \npp\ where the $n$
numbers $X_1,\ldots,X_n$ are taken to be independently and
identically distributed according to some density $\rho(X)$.  In
most cases studied so far, the $X_i$ are taken to be drawn uniformly
from a bounded domain, say integer values drawn uniformly from
$\{1,\ldots,2^m\}$ or real values drawn uniformly from $[0,1]$.  The
statistical mechanics of this model has been discussed by several
authors
\cite{fu:89,ferreira:fontanari:98,mertens:98a,sasamoto:etal:01}.

When $\rho(X)$ is the uniform distribution on $\{1,\ldots,2^m\}$,
it turns out that the typical
properties of random instances depend on the ratio $\kappa=m/n$.
Numerical simulations suggested that in the limit $n,m\to\infty$
with $\kappa$ fixed, this system had a sharp transition at
$\kappa = 1$ between a phase in which there are exponentially
many optimal solutions with energy
$0$ or $1$, and a phase where
the optimal solution is unique (except for trivial symmetry)
and has energy scaling with $2^n$ \cite{gent:walsh:96}.
This was supported by a
statistical physics approach \cite{mertens:98a} and
confirmed by rigorous analysis
\cite{borgs:chayes:pittel:01}.

For the random \npp , the costs of two partitions $\bss$ and $\bss'$
are {\em a priori} highly correlated random variables. In
\cite{mertens:00a}, one of the authors made a rather surprising
``random cost approximation,'' in which the correlations of
energies near the ground state
were neglected. Within this approximation, it is easy to
calculate the statistics of the ground state and the first
excitations. Remarkably, the results of these calculations were later
confirmed by rigorous analysis \cite{borgs:chayes:pittel:01}, which
therefore suggested that there might be a  mathematical basis for
this approximation.

Numerical simulation and heuristic arguments led to an even stronger
conjecture, namely that the statistical independence of nearby
levels is not restricted to energies close to the ground state but
extends to all fixed ``typical'' energies \cite{rem1}.
These authors also
conjectured that the overlaps corresponding to these energies are
uncorrelated.  These two claims were collectively called the {\em
local REM conjecture} \cite{rem1}, since the proposed behavior of
nearby energies was analogous to that of the random energy model
(REM) in spin glass theory \cite{derrida:81}. In this paper, we
prove the local REM conjecture for the \npp\ with a general
distribution of the $X_i$.

In physical terms, the optimal partitions of the \npp\ are precisely
analogous to the ground states of a mean-field antiferromagnetic
Ising spin system with Mattis-like couplings $J_{ij}=-X_iX_j$
defined by the Hamiltonian
\begin{equation}
  \label{eq:mattis}
  H(\bss) = E^2(\bss) = \frac{1}{n}\sum_{ij}X_iX_j\sigma_i\sigma_j
  =: -\frac{1}{n}\sum_{ij}J_{ij}\sigma_i\sigma_j\,.
\end{equation}
Similarly, the energy spectrum and overlaps of the \npp\ are
analogous to those of the mean-field antiferromagnetic Mattis spin
glass.  Our results therefore also establish the REM conjecture
for this spin glass.

\section{Statement of Results}

Let $X_1,...,X_n$ be independent random variables distributed
according to the common density function $\rho(x)$. We assume that
$X$ has finite second moment and $\rho(x)$ satisfies the bound \eq
\label{dist} \int_{-\infty}^{\infty} \rho(x)^{1+\epsilon} \;
dx < \infty \en for some $\epsilon > 0$. Note that this includes, in
particular, all bounded density functions with finite second moment.
We use the symbol $\PR_n(\cdot)$ to denote the probability with respect
to the joint probability distribution of $X_1,...,X_n$ .

As in the introduction, we represent the $2^{n}$ partitions of the
integers $\{1,..,n\}$ as spin configurations $\bss \in
\{-1,+1\}^n$, define the energy of $\bss$ as in \eqref{ene}, and
denote by $E_1,\ldots ,E_N$ the increasing spectrum of the energy values
corresponding to the $N = 2^{n-1}$ distinct partitions.
We also
denote by $\bss^{(1)},\dots,\bss^{(N)}$ the configurations
corresponding to these ordered energies.

Finally, we introduce the rescaled overlaps as follows.
Consider the random variable $r_n$ defined by the condition
$E_{r_n} <\alpha \leq E_{r_n+1}$.  For $j>i>0$, the rescaled overlap
is defined by
\begin{equation}
\label{overlap-resc} Q_{ij}= \frac 1{\sqrt n}
\sum_{s=1}^n\sigma^{(r_n+i)}_s\sigma^{(r_n+j)}_s.
\end{equation}

On the basis of both heuristic arguments and
numerical evidence, Bauke, Franz and Mertens \cite{rem1}
conjectured the following
behavior for the energy level and overlap
statistics of the \npp\ with the $X_i$
uniformly distributed in $[0,1]$:
\begin{conjecture}
\label{rem} Let  $X_1,X_2,\dots,X_n$ be i.i.d.\ random variables
distributed uniformly in $[0,1]$, let $\alpha \geq 0$ be a fixed
real number, and let $l$ be a fixed positive integer.  Define $r$
by $E_r <\alpha \leq E_{r+1}$. Then
\begin{equation}
\label{rem-uniform}
\begin{aligned}
 \sqrt{\frac{6}{\pi}} 2^{n-1} e^{-3\alpha^2/2}
 &(E_{r+1}-\alpha,
E_{r+2}-\alpha,...,E_{r+l}-\alpha)\\
& \stackrel{w}{\Ra} (w_1,w_1+w_2,...,w_1+w_2+\cdots+w_l),
\end{aligned}
\end{equation}
where $w_i$ are i.i.d.\ random variables each distributed
exponentially with mean 1, and $\stackrel{w}{\Ra}$ denotes weak
convergence as $n\to\infty$. In addition, spin configurations
corresponding to different energy levels become asymptotically
uncorrelated in the sense that, for all $j > i > 0$,
the rescaled overlap $Q_{ij}$
converges to a standard normal.
\end{conjecture}

For $\alpha=0$,  the part of the conjecture
concerning the energies was already rigorously established
in \cite{borgs:chayes:pittel:01}.
In this paper we
prove that the full Conjecture \ref{rem} for fixed $\alpha > 0$
holds not
only for the uniform distribution,
but also for any distribution
which has finite second moment and satisfies (\ref{dist}).

\begin{theorem}
\label{mainth} Let $\rho$ be a  probability density on
$[-\infty,\infty]$ with finite second moment $\tau^2$. Assume that
$\rho$ satisfies the condition \eqref{dist} for some $\epsilon > 0$,
let $l$ be a fixed positive integer, and let $\alpha$ be a fixed
real number.  If $X_1,...,X_n$ are independent random variables
distributed according to $\rho$, and $r_n$ is defined so that
$E_{r_n} <\alpha \leq E_{r_n+1}$, then
\begin{equation}
\label{rem-general}
\begin{aligned}
\sqrt{\frac{2}{\pi \tau^2}} &2^{n-1} e^{-\alpha^2/(2\tau^2)}
(E_{r_n+1}-\alpha, E_{r_n+2}-\alpha,...,E_{r_n+l}-\alpha)
\\
&\qquad\stackrel{w}{\Ra} (w_1,w_1+w_2,...,w_1+w_2+\cdots+w_l)
\end{aligned}
\end{equation}
 where $w_i$ are i.i.d.
random variables each distributed exponentially with mean 1, and
$\stackrel{w}{\Ra}$ denotes weak convergence as $n\to\infty$.
In
addition, spin configurations corresponding to different energies
become asymptotically uncorrelated in the sense that for
any fixed $j>i>0$, the rescaled overlap $Q_{ij}$ converges in
distribution to
a standard normal, i.e.,
\begin{equation}
\label{PQ-lim} \lim_{n\to\infty} \PR_n(Q_{ij}\geq \beta)=
\int_{\beta}^{\infty} \frac{1}{\sqrt{2\pi}}\, e^{-\frac{x^2}{2}} dx.
\end{equation}
\end{theorem}

\begin{remark}
\label{remarkmain} \noindent
\begin{enumerate}
\item Substituting $\tau^2 = \frac{1}{3}$ for the case when $X_i$ is
  distributed as $U[0,1]$, observe that Theorem \ref{mainth} reduces to
  Conjecture \ref{rem}.

\item  Having established the original REM Conjecture \ref{rem},
the question naturally arises whether analogous results hold for
energy scales $\alpha$ which grow with $n$.
Indeed, the authors of \cite{rem1} said that
they believe that the conjecture
might extend to values of $\alpha$ that grow slowly
enough with $n$, although computational limitations prevented them
from supporting this stronger claim by simulations.
In a second paper
\cite{BCMN-2},
we will show that, under suitable
  additional assumptions on the distribution $\rho$, the
  conjecture does indeed hold provided
  $\alpha=o(n^{1/4})$.
  \end{enumerate}
\end{remark}

In addition to immediately implying the analogous
results for the mean-field antiferromagnetic Mattis spin glass
(see equation \eqref{eq:mattis}),
our theorem on the energy spectrum of the \npp\ also gives the
energy spectrum of the one-dimensional Edwards-Anderson ($1$-d EA)
spin glass model away from the ground state.  The $1$-d EA model has
energy $E(\sigma) = \sum_i J_i \sigma_i \sigma_{i+1}$.  Consider the
transformation $\tau_i = \sigma_i \sigma_{i+1}$ and take the
boundary condition  $\sigma_{n+1} = 1$.  Then $E(\sigma) =  \sum_i
J_i \tau_i$, so that,
up to a multiplicative factor of $\sqrt n$,
the energy of the \npp\ with random variables
$X_i$ is the same as the absolute value of the energy of the the
$1$-d EA model with random variables $J_i$.  Note that the energy
spectrum of the \npp\ lies in $[0, E_{\rm max}]$, with $E_{\rm max}
= \theta(\sqrt n)$,
 while that of the
$1$-d EA model lies in $[-\sqrt{n}E_{\rm max},\sqrt{n}E_{\rm max}]$.

Our theorem says that properly scaled energies of the \npp\ converge
to a Poisson process.  By the above transformation, this result
obviously applies also to the $1$-d EA model except for energies
about zero, which are correlated by symmetry. In particular, the
result applies to the $1$-d EA model in energy intervals of the form
$[\sqrt{n}\alpha, \sqrt{n}(\alpha + \theta(e^{-n}))]$
for any bounded $\alpha \geq 0$ or their reflection about $0$.  If
the interval includes the origin as an internal point, the positive
and negative energies separately converge to Poisson processes, with
the two obviously related by a spin-flip symmetry.
\section{Proof of Theorem~\ref{mainth}}

\subsection{Outline of the Proof}

Before we proceed with our proof, note that we may assume without
loss of generality that the second moment $\tau^2$ is equal to $1$
and that $\rho$ is symmetric, $\rho(x)=\rho(-x)$. Indeed,
considering the rescaled random variables $\tilde X_i=\tau^{-1}
X_i$, we immediately see that the statements of theorem for general
$\tau$ follow from those for $\tau=1$. Next, consider the random
variables $Y_1,...,Y_n$ where each $Y_i$ is obtained as $X_i$ w.p.
$\frac{1}{2}$ or $-X_i$ w.p. $\frac{1}{2}$. It is easy to see that
the energy spectrum of the $Y_1,...,Y_n$ is identical to that of the
$X_1,...,X_n$. Further, from convexity of $\abs{x}^{1+\epsilon}$ and
Jensen's inequality, it follows that $\rho_Y(y)=\frac
12(\rho(x)+\rho(-x))$ also satisfies (\ref{dist}). Therefore,
w.l.o.g.\ we can assume that $\rho(x) = \rho(-x)$ as claimed,
and in particular that $X_i$ has zero first moment.
For simplicity of notation, we omit the subscript $n$, and
 denote the probability with respect to the joint distribution
of $X_1,\dots,X_n$ by $\PR(\cdot)$, and the expectation with respect
to this distribution by $\EX(\cdot)$.

Let $Z_n(a,b)$ be the number of points of the energy spectrum that
lie in the interval $[a,b]$, and let $N_n(t)$ be the number of
points in the energy spectrum that fall into the (shifted and)
re-scaled interval $[\alpha,\alpha + t\xi_n]$, where
\eq
\xi_n=\sqrt{\frac{\pi }{2}} 2^{-(n-1)} e^{\alpha^2/2}.
\en
 We must
show that $N_n(t)$ converges to a Poisson process with parameter
one.  To this end, we will show that for any family of
non-overlapping intervals $[c_1,d_1]$, $\dots$,
$[c_m,d_m]$ with $d_i> c_i\geq 0$, the rescaled
variables $Z_n(a_n^i,b_n^i)$ with
$a_n^i=\alpha + c_i \xi_n$ and $b_n^i=\alpha + d_i \xi_n$
converge  in distribution to the {\em
increments} of a Poisson process with parameter one. We prove this
by showing the convergence of the multidimensional factorial
moments, i.e., by proving
the following theorem.

\begin{theorem}
\label{main-thm}
Let $\alpha\geq 0$,
let $m$ be a positive integer, and
let $[c_1,d_1]$, $\dots$,
$[c_m,d_m]$ be a family of non-overlapping intervals.
For $\ell=1,\dots,m$,
set $a_n^\ell=\alpha + c_\ell \xi_n$ and $b_n^\ell=\alpha + d_\ell\xi_n$,
where
$\xi_n=\sqrt{\frac{\pi }{2}} 2^{-(n-1)} e^{\alpha^2/2}$.
Given an arbitrary $m$-tuple
$(k_1,..,k_m)$ of positive integers, we then have
\eq
\label{mulfact} \lim_{n\ra\infty}
\mathbb{E}[\prod_{\ell=1}^m (Z_n(a_n^{\ell},b_n^{\ell}))_{k_{\ell}}] =
\prod_{\ell=1}^m
(d_\ell-c_\ell)^{k_{\ell}},
\en
 where, as usual, $(Z)_k=Z(Z-1)\dots (Z-k+1)$.
\end{theorem}

Theorem~\ref{main-thm} establishes that  $N_n(t)$ converges to a Poisson
with rate one.   As we will see
below, the asymptotic independence of configurations corresponding
to nearby energy levels is an immediate corollary to the proof of
this theorem.

\subsection{Integral Representation}
Following the general strategy employed in Section 6 of
\cite{borgs:chayes:pittel:01}, we base
the proof of Theorem~\ref{main-thm} on an integral
representation of the factorial moments.  The derivation of
this integral representation
 uses the Fourier transform of
$\rect(x)$, where as usual $\rect(x)$ is defined by
\eq
\label{rect}
\rect(x) = \left\{ \begin{array}{ll} 1 & -\frac{1}{2} \leq x \leq
  \frac{1}{2}\\
0 & \text{otherwise}.\end{array} \right.
\en
Let $t_n^{\ell} = \frac
{a_n^{\ell}+b_n^{\ell}}{2}$ denote the center of the interval
$[a_n^{\ell},b_n^{\ell}]$, let $\gamma_\ell=d_\ell-c_\ell$,
and let $q_{n,\ell}=\gamma_\ell\xi_n\sqrt{n}$.  Then
$Z_n(a_n^{\ell},b_n^{\ell})$ can be written as
\eq
\label{Z1} Z_n(a_n^{\ell},b_n^{\ell}) =
 \sum_{\bss} I^{(\ell)}(\bss)
\en
where
\eq
\label{I1}
 I^{(\ell)}(\bss)
=
\frac{1}{2} \left[ \rect(\frac{\sum_{s=1}^n \sigma_s X_s
-t_n^{\ell}\sqrt{n}}{q_{n,\ell}}) + \rect(\frac{\sum_{s=1}^n
\sigma_s X_s +
    t_n^{\ell}\sqrt{n}}{q_{n,\ell}})  \right].
\en
Note  the factor $\frac{1}{2}$, which arises from the fact that each
partition is counted only once in $Z_n(a_n^{\ell},b_n^{\ell})$, while the
two
configurations $\bss$ and  $-\bss$ correspond to the same partition
of $\{1,\dots,n\}$.

Next we write the
 ${k}^\text{th}$
factorial moment of
$Z_n(a_n^{\ell},b_n^{\ell})$
as a sum over sequences of $k$
distinct configurations $\bss^{(1)},\dots, \bss^{(k)}$.
To this end, let us first note
that $I^{(\ell)}(\bss)$ is either $0$ or $1/2$,
implying that
\begin{equation}
I^{(\ell)}(\bss)=I^{(\ell)}(\bss)(I^{(\ell)}(\bss)+I^{(\ell)}(-\bss))
\end{equation}
for all $\bss\in\{-1,+1\}^n$.  Using this fact, we now rewrite the
$k^\text{th}$
factorial moment as
\begin{equation}
\label{Zm} \mathbb E[( Z_n(a_n^{\ell},b_n^{\ell}))_{k} ]
= \sum_{\pm\bss^{(1)} \neq \cdots \neq
  \pm \bss^{(k)}}
  \mathbb E[I^{(\ell)}_{k}(\bss^{(1)},\dots, \bss^{(k)})]
\end{equation}
where the sum runs over distinct configurations and
\begin{equation}
\label{Im}
I_{k}^{(\ell)}(\bss^{(1)},\dots, \bss^{(k)})=
  \prod_{j=1}^{k} I^{(\ell)}(\bss^{(j)}),
\end{equation}
with $I^{(\ell)}(\cdot)$ given by \eqref{I1}.

To obtain a formula for the multi-dimensional factorial
moments, let us consider two disjoint intervals
$[a_n^{\ell},b_n^{\ell}]$ and $[a_n^{\ell'},b_n^{\ell'}]$, and two sequences
of
configurations $\bss^{(1)},\dots,\bss^{(k_{\ell})}$ and
$\tilde\bss^{(1)},\dots,\tilde\bss^{(k_{\ell'})}$
contributing to $(Z_n(a_n^{\ell},b_n^{\ell}))_{k_{\ell}}$ and
$(Z_n(a_n^{\ell'},b_n^{\ell'}))_{k_{\ell'}}$, respectively.
Recall that the energy of the configuration $\bss$ is equal to
the energy of the configuration $-\bss$, $E(\bss)=E(-\bss)$.
Since $I^{(\ell)}(\bss)=0$ unless
$E(\bss)\in [a_n^{\ell},b_n^{\ell}]$
and $ I^{({\ell'})}(\tilde\bss)=0$ unless
$E(\tilde\bss)\in [a_n^{\ell'},b_n^{\ell'}]$, we see that
$I^{(\ell)}(\bss)I^{({\ell'})}(\tilde\bss)=0$
if $\bss$ and $\tilde\bss$ are not distinct.
The combined sequence
$
\bss^{(1)},\dots,\bss^{(k_{\ell})},
\tilde\bss^{(1)},\dots,\tilde\bss^{(k_{\ell'})}
$
therefore only contributes to the product
$(Z_n(a_n^{\ell},b_n^{\ell}))_{k_{\ell}}
(Z_n(a_n^{\ell'},b_n^{\ell'}))_{k_{\ell'}}$ if
$\bss^{(j)}\neq \pm \tilde\bss^{({\ell'})}$
for all ${\ell}\neq{\ell'}$.  As a consequence, the
multi-dimensional factorial
moment in Theorem~\ref{main-thm} is itself given as sum over sequences
of pairwise distinct configurations.
More explicitly, let
$k=\sum_{\ell=1}^m k_{\ell}$, and for $j=1,\dots, k$,
let
$\ell(j)=1$ if $j=1,\dots,k_1$, $\ell(j)=2$ if $j=k_1+1,\dots, k_1+k_2$,
and so on.  Then
\begin{equation}
\label{Zm-multi}
\mathbb{E}[\prod_{\ell=1}^m (Z_n(a_n^{\ell},b_n^{\ell}))_{k_{\ell}}]
=\sum_{\pm\bss^{(1)} \neq \cdots \neq
  \pm \bss^{(k)}}
  \mathbb E[I_k(\bss^{(1)},\dots, \bss^{(k)})]
\end{equation}
where
\begin{equation}
\label{Im-multi}
I_{k}(\bss^{(1)},\dots, \bss^{(k)})=
  \prod_{j=1}^{k} I^{(\ell(j))}(\bss^{(j)}).
\end{equation}

Using the Fourier inversion theorem and the fact that the Fourier
transform of the function $\rect(x)$ is equal to
\[ \sinc(f) =  \frac{\sin \pi f}{\pi f} , \]
we then rewrite
$I^{(\ell)}(\bss)$ as
\eq \label{Iell-int3}
I^{(\ell)}(\bss) = q_{n,\ell}
\int_{-\infty}^{\infty}
 \;
\sinc(f q_{n,\ell} )\cos(2\pi f t_n^{\ell}\sqrt{n}) e^{2\pi i f \sum_{s=1}^n
\sigma_s X_s} df,
\en
leading to the representation
\eq \label{Z3}
\mathbb E[Z_n(a_n^{\ell},b_n^{\ell})] = q_{n,\ell}
\sum_{\mb{\sigma}}
\mathbb E\Bigl[\int_{-\infty}^{\infty}
 \;
\sinc(f q_{n,\ell} )\cos(2\pi f t_n^\ell \sqrt{n}) e^{2\pi i f \sum_{s=1}^n
\sigma_s X_s} df\Bigr]
\en
and a similar representation for the  factorial
moments.  As we will see, the expectation and the integral in
\eqref{Z3} can be interchanged, leading to a representation of the
first moment, $\mathbb E[Z_n(a_n^{\ell},b_n^{\ell})]$, in terms of the
 Fourier transform
\eq
\label{unihyp} \rhat(f)=\mathbb{E}[e^{2\pi i f X}].
\en
In a similar way, the factorial moments can be expressed
in terms of the Fourier transform $\rhat$ of of the distribution
function $\rho$.

We will use several properties of the Fourier transform in
our proof, which we summarize now.
All of them follow from the fact
that the density $\rho(x)$  has finite second moment, satisfies
equation (\ref{dist}), and is symmetric under the transformation
$x\to -x$.
\begin{enumerate}
\item[$(i)$]
\label{rhat-Prop1} For any $\mu_1 > 0$, there exists $c_1 >0$,
possibly
  depending on $\mu_1$, such that whenever $|f| \geq \mu_1$, we
  have $\abs{\rhat(f)} < e^{-c_1}$.
\item[$(ii)$]
\label{rhat-Prop2} For any $n \geq n_o$, where $n_0$ is the solution
of $\frac{1}{1+\epsilon}
+ \frac{1}{n_o} =
  1$ with $\epsilon$ as in \eqref{dist}, we have
\begin{equation}
\label{bdrhat}
 \int_{-\infty}^{\infty} |\rhat(f)|^{n} \leq \int_{-\infty}^{\infty}
|\rhat(f)|^{n_0} = C_0 < \infty.
\end{equation}
\item[$(iii)$]
\label{rhat-Prop3} $\rhat(f) \ra 0$ as $|f| \ra \infty$ .
\item[$(iv)$] There exists a $c_2 > 0$ such that, for $\mu_1 >0$ small
enough,
whenever $|f| \leq \mu_1$, we have
$|\rhat(f)|  \leq e^{-c_2f^2}$.
\end{enumerate}

\subsection{First Moment}
\label{section:firstmoment}

In this subsection, we calculate the first moment of
the random variable
$Z_n(a_n^{\ell},b_n^{\ell})$.
To avoid very cumbersome notation, we omit the index $\ell$ in this
subsection,
and write
 $a_n$, $b_n$, $\gamma$ and $q_n$ for $a_n^{\ell}$, $b_n^{\ell}$,
$\gamma_\ell$
 and $q_{n,\ell}$,
 respectively.

We first show that we can
exchange the expectation with respect to $\rho$ with
the integral in \eqref{Z3}.  This is the content of the following lemma.

\begin{lemma}
\label{lem:lim-exch} For all $n\geq 1$,
\begin{equation}
\label{ezn}
\mathbb{E}[Z_n(a_n,b_n)] = 2^n q_n\int_{-\infty}^{\infty}
\; \sinc(fq_n)\cos(2\pi f t_n\sqrt{n})\rhat_n(f) df .
\end{equation}

\end{lemma}

\begin{proof}
We use truncation to justify the interchange of the integral and the
expectation in equation \eqref{Z3}. For any $B>0$, define \eq
\label{trunc} Z_n^{(\leq B)}(a_n,b_n) = q_n\sum_{\bss} \int_{-B}^{B}
\; \sinc(fq_n)\cos(2\pi f t_n\sqrt{n})e^{2\pi i f \sum_{s=1}^n
\sigma_s X_s} df .\en Observe that this corresponds to computing the
inverse Fourier transform after truncating the Fourier transform in
the region $[-B,B]$. Thus we can write
\eq \label{Zinv} Z_n^{(\leq B)}(a_n,b_n) = \frac{1}{2} \sum_{\bss}
f_B(\frac{\sum_{s=1}^n \sigma_s
  X_s -t_n\sqrt{n}}{q_n}) + f_B(\frac{\sum_{s=1}^n \sigma_s X_s +
  t_n\sqrt{n}}{q_n}) ,
\en
where $f_B(x)$ is the appropriately defined inverse Fourier
transform for the truncated integral. It is known that $f_B(x) \ra
\rect(x)$ as $B \ra \infty$ for all $x \neq \pm \frac{1}{2}$. At $x
= \pm \frac{1}{2}$, $f_B(x) \ra \frac{1}{2}$. Since $Z_n^{(\leq
B)}(a_n,b_n)$ is a finite sum we see that $Z_n^{(\leq
B)}(a_n,b_n)$ converges almost surely to $Z_n(a_n,b_n)$. The bounded
convergence theorem then implies
\eq \label{bd1}
\mathbb{E}[Z_n(a_n,b_n)]
= \lim_{B \ra \infty}\mathbb{E}[Z_n^{(\leq B)}(a_n,b_n)] .
\en

Using Fubini's theorem, independence of the $X_i$ and the fact that
$\sigma_i X_i$ has the same distribution as $X_i$, we may express
the expectation of the right hand side of \eqref{trunc} as
\begin{equation}
\label{trun2}
\begin{aligned}
\mathbb{E}[Z_n^{(\leq B)}(a_n,b_n)] &= 2^nq_n \int_{-B}^{B} \;
\sinc(f q_n)\cos(2\pi  f t_n\sqrt{n}) \mathbb{E}[e^{2\pi i f X}]^n
df
\\
&=
 2^nq_n \int_{-B}^{B} \;
\sinc(fq_n) \cos(2\pi f t_n\sqrt{n})\rhat_n(f) df .
\end{aligned}
\end{equation}

We now combine H\"older's inequality with the bound \eqref{bdrhat}
and the fact that $\sinc(f q_n)$ is in $L^{1+\epsilon}$ for all
$\epsilon>0$ to bound \eq \label{trun2a}
\begin{aligned}
\int_{-B}^{B} \; |\sinc&(fq_n) \cos(2\pi f t_n\sqrt{n})\rhat_n(f)
|df \leq \int_{-\infty}^{\infty} \; |\sinc(fq_n)| |\rhat(f) |df
\\
&\leq \Bigl(\int_{-\infty}^{\infty} \; |\sinc(fq_n)|^{1+\epsilon}
df\Bigr)^{1/(1+\epsilon)}\; \Bigl(\int_{-\infty}^{\infty} \;
 |\rhat(f) |^{n_0}df\Bigr)^{1/n_0}<\infty.
\end{aligned}
\en By dominated convergence, the integral in (\ref{trun2})
therefore converges as $B \ra \infty$, giving \eq \label{fineq}
\begin{split}
\mathbb{E}[Z_n(a_n,b_n)] &= \lim_{B \ra \infty}
\mathbb{E}[Z_n^{(\leq B)}(a_n,b_n)]\\ &=
2^nq_n\int_{-\infty}^{\infty}  \; \sinc(fq_n) \cos(2\pi f t_n
\sqrt{n})\rhat_n(f) df,
\end{split}
\en as required.
\end{proof}

Having established the integral representation \eqref{ezn} for
 ${\mathbb E}[Z_n(a_n,b_n)]$, we are now ready to prove the
convergence of the first moment.

\begin{lemma}
\label{lem:Z-asymp}
\begin{equation}
\label{fineq1}
\lim_{n\to\infty} \mathbb{E}[Z_n(a_n,b_n)]= \gamma.
\end{equation}
\end{lemma}

\begin{proof}
The proof is a standard application of well-known techniques from
asymptotic analysis.
However, for the convenience of the reader, and due to the
fact that our later proof for higher factorial moments relies
on some of the techniques used here, we present the full details.

The proof proceeds in two steps.  In the first step we show
that, at the cost of an error which is negligible as $n\to\infty$,
the integral in \eqref{ezn}
can be restricted to a small neighborhood of $0$ with the
$\sinc$ factor replaced
by $1$, and in the second step we expand $\rho^n(f)$ around $0$
to get a Gaussian approximation for the integral.

We first show that, given any sequence
$\omega_n$  with
$\omega_n\to\infty$ and $\omega_n/\sqrt n\to 0$
as $n\to\infty$, we have

\begin{equation}
\label{limapp1}
\mathbb{E}[Z_n(a_n,b_n)] = 2^n q_n \int_{-\omega_n/\sqrt n}^{\omega_n/\sqrt
n}
\; \cos(2\pi f \alpha \sqrt{n})\rhat_n(f) df +o(1).
\end{equation}

To this end, let us first restrict the integral in
\eqref{ezn} to $|f|\leq \mu_1$.
Since $|\rhat_n(f)| \leq
|\rhat^{n_0}(f)|e^{-c_1(n-n_0)}$ when $n>n_0$ and $|f| > \mu_1$, the
contribution from $\abs{f} > \mu_1$ to the integral on the right
hand side of \eqref{ezn} is exponentially small in $n$
 for large  $n$. In the interval $\abs{f} \leq \mu_1$,
we can replace $\sinc(fq_n)$ by $1+O((fq_n)^2)$
as $n\to\infty$, since
$q_n\to 0$ as $n\to\infty$.
Observing that
$t_n\sqrt n=\alpha\sqrt n+O(\sqrt n \xi_n)=\alpha\sqrt n+O(q_n)$,
we can also replace
$\cos(2\pi f t_n \sqrt{n})$
by $\cos(2\pi f \alpha \sqrt{n})+ O(fq_n)$.
Replacing $\sinc(fq_n)$ by $1$
and $\cos(2\pi f t_n \sqrt{n})$
by $\cos(2\pi f \alpha \sqrt{n})$, we thus
incur an error that can be bounded by $2^n O(q_n^2)=o(1)$, giving
the bound
\eq \label{fineq2} \mathbb{E}[Z_n(a_n,b_n)] =
2^n q_n \int_{-\mu_1}^{\mu_1} \; \cos(2\pi f \alpha
\sqrt{n})\rhat_n(f) df +o(1).
\en
To complete the proof of \eqref{limapp1}, we have to show that
\[
\lim_{n\to\infty} 2^nq_n\int_{\mu_1>\abs{f} > \frac{\omega_n}{\sqrt{n}}}
\; \cos(2\pi f \alpha \sqrt{n})\rhat_n(f) df = 0  .\]
Recall by property {\em (iv)} following \eqref{bdrhat} that
$\rhat (f) \leq e^{-c_2f^2}$, which implies
\[
\Bigl|
  \int_{\mu_1 > \abs{f} >  \frac{\omega_n}{\sqrt{n}}}
   \cos(2\pi f \alpha \sqrt{n})\rhat_n(f) df
\Bigr| \leq
 \int_{ \abs{f} >  \frac{\omega_n}{\sqrt{n}}}
 e^{-c_2nf^2} df
 =O(\frac 1{\sqrt n}e^{-c_2\omega^2_n}).
\]
Since $\frac 1{\sqrt n}2^nq_ne^{-c_2\omega_n^2}\to 0$ as
$n\to\infty$, this completes
the proof of \eqref{limapp1}.

To evaluate the integral in \eqref{limapp1}, we
make a Gaussian approximation in a neighborhood near origin.
While this is standard if one assume that the density
$\rho$ has a sufficiently high moment (anything more than the
second moment is enough), a little care is needed due to the fact
that we only assume existence of the second moment.

We start by choosing the sequence $\omega_n$.  Since $\rho$ has
a finite second moment, its Fourier transform is
twice continuously differentiable, implying that
$\rhat(f)= 1-2\pi^2(1+o(1)) f^2= e^{-2\pi^2 f^2(1+o(1))}$.
In other words, for $\mu_1$ sufficiently small and
$|f|\leq \mu_1$,
we can write
$\rhat(f)$ in the form
$\rhat(f)=e^{-2\pi^2 f^2+g(f)f^2}$
where $g(f)\to 0$ as $f\to 0$.  Choose a sequence which goes to zero
as $n\to\infty$, say $\log n/\sqrt n$.  We then define
\[
\epsilon_n=\sup_{f:|f|\leq \log n/\sqrt n}|g(f)|
\qquad\text{and}\qquad
\omega_n=\min\{\log n,\epsilon_n^{-1/3}\}.
\]
For $|f|\leq \omega_n/\sqrt n$, we then have
$|nf^2 g(f)|\leq \omega_n^2\epsilon_n\leq \epsilon_n^{1/3}$,
implying that
$\rhat_n(f)=e^{-2\pi^2 n f^2+o(1)}=e^{-2\pi^2 n f^2}(1+o(1))$.
 As a
consequence,
\begin{equation}
\label{large-n-asympt}
\begin{aligned}
& 2^n q_n\int_{\abs{f} < {\frac{\omega_n}{\sqrt{n}}}}
\; \cos(2\pi f \alpha \sqrt{n})\rhat_n(f) df
 \\
&\quad  = 2^nq_n \int_{\abs{f} < \frac{\omega_n}{\sqrt{n}}}\;
\cos(2\pi f \alpha \sqrt{n})e^{-2 \pi^2 n f^2} df
+o(2^nq_n )\int_{\abs{f} < \frac{\omega_n}{\sqrt{n}}}\;
e^{-2 \pi^2 n f^2} df
\\
&\quad  = 2^nq_n \int_{\abs{f} < \frac{\omega_n}{\sqrt{n}}}\;
\cos(2\pi f \alpha \sqrt{n})e^{-2 \pi^2 n f^2} df
+ o({2^n
q_n}{n^{-1/2}})
\\
&\quad  = 2^nq_n \int_{\abs{f} < \frac{\omega_n}{\sqrt{n}}}\;
\cos(2\pi f \alpha \sqrt{n})e^{-2 \pi^2 n f^2} df + o(1) .
\end{aligned}
\end{equation}

By an argument similar to proof of \eqref{limapp1}, one can show
that
\[ \lim_{n\to\infty} \int_{\abs{f} > {\frac{\omega_n}{\sqrt n}}}2^nq_n
\; \cos(2\pi f \alpha\sqrt{n})e^{-2 \pi^2 n f^2} df = 0 ,\]
implying that
\begin{equation}\begin{aligned}
\label{fineq3} \lim_{n\to\infty} \mathbb{E}[Z_n(a_n,b_n)] & =
\lim_{n\to\infty} 2^nq_n \int_{-\infty}^{\infty}
\; \cos(2\pi f \alpha\sqrt{n})e^{-2 \pi^2 n f^2}  df \\
& = \lim_{n\to\infty}
2^nq_n \frac 1{\sqrt{2\pi n}}e^{-\frac12\alpha^2}
=\gamma.
\end{aligned}\end{equation}
This completes the proof of
equation~(\ref{limapp1}).
\end{proof}

\subsection{Higher Moments}

In this subsection, we analyze the higher moments.
Bearing in mind the similar structure of the representations
\eqref{Zm} and \eqref{Zm-multi}, we first consider the
one-dimensional factorial moments $\mathbb E[(
Z_n(a_n^{\ell},b_n^{\ell}))_{k} ]$.
As in Subsection~\ref{section:firstmoment}, we omit the index $\ell$,
and write
 $a_n$, $b_n$, $\gamma$ and $q_n$ for $a_n^{\ell}$, $b_n^{\ell}$,
$\gamma_\ell$
 and $q_{n,\ell}$,
 respectively.  We also write $I(\cdot)$ and $I_{k}(\cdot)$
 instead of $I^{(\ell)}(\cdot)$ and $I_{k}^{(\ell)}(\cdot)$.

We want to show that in the limit $n\to\infty$,
the factorial moment
$\mathbb E[( Z_n(a_n,b_n))_k]$ is equal to $\gamma^k$.
Since the sum in \eqref{Zm} contains
$(2^n)_k=2^{nk}(1+o(1))$ terms, one might therefore try to
show that  $\mathbb E[I_k(\bss^{(1)},\dots, \bss^{(k)})]$
is asymptotically equal to $\gamma^k 2^{-nk}$
by generalizing our approach from the last section, which
showed that $\mathbb E[I(\bss^{(1)})]$ is asymptotically
equal to $\gamma 2^{-n}$.

But in contrast to the expectation of $I(\bss^{(1)})$,
the expectations of the random variables
$I_k(\bss^{(1)},\dots, \bss^{(k)})$ cannot be analyzed
easily for all sequences of distinct configurations
$\bss^{(1)},\dots, \bss^{(k)}$.  This problem already
appeared in \cite{borgs:chayes:pittel:01}, but here it will
be harder to overcome.  First of all, even the analog
of Lemma~\ref{lem:lim-exch} for the
higher moments will not hold unless the configurations
$\bss^{(1)},\dots, \bss^{(k)}$ form a set of $k$ linearly independent
vectors in $\mathbb R^n$.  But more importantly, the expectation of
$I_k(\bss^{(1)},\dots, \bss^{(k)})$ will be hard to analyze, even
if $\bss^{(1)},\dots, \bss^{(k)}$  are linearly independent, unless
we impose additional conditions on the configurations
$\bss^{(1)},\dots, \bss^{(k)}$.

To overcome these problems,
we use the following strategy: first we prove the analog of
Lemma~\ref{lem:lim-exch} for
$\mathbb E[I_k(\bss^{(1)},\dots, \bss^{(k)})]$ under the assumption that
the configurations
$\bss^{(1)},\dots, \bss^{(k)}$  are linearly independent.
Then we formulate  a condition on the
sequence $\bss^{(1)},\dots, \bss^{(k)}$
that allows
us to analyze  $\mathbb E[I_k(\bss^{(1)},\dots, \bss^{(k)})]$
by an extension of the proof of Lemma~\ref{lem:Z-asymp}.  Having extracted
the leading behavior, we then estimate
the contributions of all other configurations
$\bss^{(1)},\dots, \bss^{(k)}$  and show that they
do not contribute in the limit $n\to\infty$.

We start with the analog of
 Lemma~\ref{lem:lim-exch} for the
higher moments.

\begin{lemma}
\label{lem:int-exch-u} Let ${k}$ be a positive integer,
and let $\bss^{(1)},..,\bss^{({k})}$
be linearly independent configurations in $\{-1,+1\}^n$. Then
\begin{equation}
\label{int3} \mathbb{E}[I_{k}(\bss^{(1)},\dots, \bss^{({k})})] =
q_n^{k}\iiint_{-\infty}^{\infty} \;  \prod_{s=1}^n \rhat(v_s)
\prod_{j=1}^{k} \sinc(f_j q_n)\cos(2\pi f_j t_n \sqrt{n})  df_j,
\end{equation}
where
\begin{equation}
\label{vs} v_s = \sum_{j=1}^{k} \sigma_s^{(j)} f_j, \qquad 1 \leq s \leq n.
\end{equation}

\end{lemma}
\begin{proof}
Let us first rewrite $I_{k}(\bss^{(1)},\dots, \bss^{({k})})$ as
\begin{equation}
\label{int1}
\begin{aligned}
& I_{k}(\bss^{(1)},\dots, \bss^{({k})})\\
& \quad = q_n^{k}\iiint_{-\infty}^{\infty} \; \prod_{s=1}^n e^{2\pi i
X_s \sum_{j=1}^{k} \sigma_s^{(j)} f_j} \prod_{j=1}^{k} \sinc(f_j
q_n)\cos(2\pi f_j t_n \sqrt{n}) df_j
 \\
&  \quad= q_n^{k}\iiint_{-\infty}^{\infty} \; \prod_{s=1}^n e^{2\pi i
X_s v_s} \prod_{j=1}^{k} \sinc(f_j q_n)\cos(2\pi f_j t_n \sqrt{n})
df_j.
\end{aligned}
\end{equation}
Arguing as in the proof of Lemma~\ref{lem:lim-exch}, we then have
\begin{equation}
\mathbb{E}[I_{k}(\bss^{(1)},\dots, \bss^{({k})})]
=\lim_{B\to\infty}\mathbb{E}[ I_{k}^{(\leq B)}(\bss^{(1)},\dots,
\bss^{({k})})],
\end{equation}
where
\begin{equation}
\label{int1m}
\begin{aligned}
& I_{k}^{(\leq B)}(\bss^{(1)},\dots, \bss^{({k})}) =
q_n^{k}\iiint_{-B}^{B} \; \prod_{s=1}^n e^{2\pi i X_s v_s}
\prod_{j=1}^{k} \sinc(f_j q_n)\cos(2\pi f_j t_n \sqrt{n}) df_j.
\end{aligned}
\end{equation}
Using again Fubini's theorem and independence of the $X_i$, we now
get \eq \label{trun2-high}
\begin{aligned}
\mathbb{E}[ I_{k}^{(\leq B)}(\bss^{(1)},\dots, \bss^{({k})})] &=
q_n^{k}\iiint_{-B}^{B} \;  \prod_{s=1}^n \rhat(v_s) \prod_{j=1}^{k}
\sinc(f_j q_n)\cos(2\pi f_j t_n \sqrt{n})  df_j.
\end{aligned}
\en To continue, we will use the fact that
$\bss^{(1)}\dots,\bss^{({k})}$ are linearly independent, implying that
the matrix $M_{k}=[\sigma_s^{(j)}]_{j \leq {k}, s \leq n}$ has rank ${k}$.
Relabeling, if necessary, let us assume that
$[\sigma_1^{(j)}]_{j\leq {k}},\dots,[\sigma_{k}^{(j)}]_{j\leq {k}}$ form a
basis of the row space.  H\"older's inequality, the fact that
$|\rhat(v_s)|\leq 1$, and a change of variables from $f_1,\dots,f_{k}$
to $v_1,\dots,v_{k}$
 then
leads to the bound \eq \label{trun2-highm}
\begin{aligned}
&\iiint_{-B}^{B} \;  \Bigl|\prod_{s=1}^n \rhat(v_s) \prod_{j=1}^{k}
\sinc(f_j q_n)\cos(2\pi f_j t_n \sqrt{n})\Bigr|  df_j
\\
&\qquad\leq \Bigl(\iiint_{-\infty}^{\infty} \; \prod_{j=1}^{k}|
\sinc(f_j q_n)|^{1+\epsilon}  df_j\Bigr)^{1/(1+\epsilon)}
\Bigl(\iiint_{-\infty}^{\infty} \; \Bigl|\prod_{s=1}^{k}
|\rhat(v_s)|^{n_0} \prod_{j=1}^{k}  df_j\Bigr)^{1/n_0}
\\
&\qquad= \Bigl(\iiint_{-\infty}^{\infty} \; \prod_{j=1}^{k}| \sinc(f_j
q_n)|^{1+\epsilon}  df_j\Bigr)^{1/(1+\epsilon)}
\Bigl(J_{k}\iiint_{-\infty}^{\infty} \; \Bigl|\prod_{s=1}^{k}
|\rhat(v_s)|^{n_0} dv_s\Bigr)^{1/n_0}
\\
&\qquad<\infty,
\end{aligned}
\en where $n_0$ and $\epsilon$ are as in the proof of
Lemma~\ref{lem:lim-exch} and $J_{k}$ is the Jacobian of the change of
variables from $f_1,\dots,f_{k}$ to $v_1,\dots,v_{k}$. By dominated
convergence, we can therefore take the limit $B\to\infty$ in
\eqref{trun2-highm}.  Putting everything together, this  gives
\eqref{int3}.
\end{proof}

Next we would like to
prove that for a ``typical set of configurations''
$\bss^{(1)},...,\bss^{({k})}$,
the integral
on the right hand side of \eqref{int3}
is equal to
$( 2\pi n)^{-{k}/2}e^{-{k}\alpha^2/2}(1+o(1))$.
Here the meaning of typical is best formulated in terms of
the
matrix
formed by the row vectors $\bss^{(1)},\dots,\bss^{(k)}$.
More generally, for $u\leq k$ and
$\bss^{(1)},\dots,\bss^{(u)}\in\{-1,+1\}^n$,
let $M_u$ be the matrix with matrix elements
$\sigma_s^{(j)}$, where $1\leq j,s\leq u$.
Given this matrix and a vector
$\bsd \in \{-1,1\}^u$, let
\begin{equation}
n_\bsd = n_\bsd(\bss^{(1)},...,\bss^{(u)}) = |
\{j\leq n: (\sigma_j^{(1)},...,\sigma_j^{(u))} = \bsd\}|
\end{equation}
 be the
number of times the column vector $\bsd$ appears in the matrix $M_u$.

If one were to choose configurations
$\bss^{(1)},\dots,\bss^{(u)}\in\{-1,+1\}^n$ independently and uniformly at
random, then for all $\delta\in\{-1,+1\}^u$,
the expectation of $n_\bsd$
is clearly equal to $n2^{-u}$.  By a standard Martingale argument,
for most configurations,
the difference  between $n_\bsd$ and $n2^{-u}$ is then not
much larger than $\sqrt n$, see
Lemma~\ref{lem:comb-bds} below.
Let us therefore assume for the moment that
\eq
\label{const-lan}
\max_{\bsd} | n_{\bsd}(\bss^{(1)},...,\bss^{(u)}) -
\frac{n}{2^u}| \leq \sqrt{n} \lambda_n
\en
for some $\lambda_n\to\infty$ to be chosen later.
The next lemma shows that under this
condition, the right hand side of \eqref{int3} behaves as desired,
provided $\lambda_n$ is chosen appropriately.
For concreteness, we chose $\lambda_n=\log n$, even though
the proof works for much larger class of sequences.

\begin{lemma}
\label{lem:Zkk-asymp}
Let
$\lambda_n=\log n$,
let $k$ be a positive integer,
and let
$\bss^{(1)},\dots,\bss^{(k)}$ be a sequence of configurations of rank
$k$ that satisfies  \eqref{const-lan}.
  Then
\begin{equation}
\label{Ik-asympt}
2^{nk}\mathbb{E}[I_k(\bss^{(1)},\dots, \bss^{(k)})]
=\gamma^{k}+ o(1),
\end{equation}
where the constant implicit in the $o$-symbol depends on $k$.
\end{lemma}

\begin{proof}
In view of Lemma~\ref{lem:int-exch-u} we will have to estimate the
expression
\begin{equation}
\label{onetwo} 2^{n{k}}q_n^{k}\iiint_{-\infty}^{\infty} \;  \prod_{s=1}^n
\rhat(v_s) \prod_{j=1}^{k} \sinc(f_j q_n)\cos(2\pi f_j t_n \sqrt{n})
df_j.
\end{equation}
Let $\mu_1$, $c_1$, $c_2$ and  $\omega_n$ be as in the proof of
Lemma \ref{lem:Z-asymp}.  In a first step, we want to show
that
the contribution of the region where
$|v_s|>\mu_1$ for at least one $s$ is negligible.

Thus consider the event that one of the $|v_s|$'s, say $|v_{t_1}|$,
is larger than $\mu_1$. Let $\bsd^1 =
\{ \sigma_{t_1}^{(1)},..,\sigma_{t_1}^{({k})} \}$, and let
$\bsd^2,...,\bsd^{k}$
be vectors such that the rank of $\{\bsd^1,..,\bsd^{k}\}$ is  ${k}$.
Let $\{v_{t_2},...,v_{t_{k}}\}$ be defined by
\[v_{t_i} = \sum_{j=1}^{k} \delta_{k}^j f_j \]
Since the vectors $\{\bsd^1,..,\bsd^{k}\}$ have rank ${k}$, we can
change the variables of integration from $f_j$ to $v_{t_j}$. Let the
Jacobian of this transformation be $J_{k}$. The Jacobian
$J_{k}$ is bounded above by
the largest determinant, $J_{max}$, of a matrix of size ${k}$ whose
entries are $\pm 1$.
We now bound the integral over the region
where $|v_{t_1}|>\mu_1$  as follows:
\begin{equation}
\label{onetwo-1}
\begin{aligned}
|I_3| &= \left| \iiint_{-\infty}^{\infty} \int_{\abs{v_{t_1}} >
\mu_1} \; J_{k} \prod_{s=1}^n \rhat(v_s)
\prod_{j=1}^{k} \sinc(f_j q_n)\cos(2\pi f_j t_n \sqrt{n})  dv_{t_j}
\right|  \\
& \leq
J_{max}\iiint_{-\infty}^{\infty}
\prod_{j=2}^{k} |\rhat(v_{t_j})|^{n_{\bsd^j}} dv_{t_j}
\times \int_{\abs{v_{t_1}}
>\mu_1} |\rhat(v_{t_1})|^{n_{\bsd^1}} dv_{t_1} \\
& \leq J_{max}(C_0)^{{k}-1} \int_{\abs{v_{t_1}} > \mu_1}
|\rhat(v_{t_1})|^{n_{\bsd^1}} dv_{t_1}
\leq J_{max} C_0^{k} e^{-c_1(n_{\bsd^1}-n_0)}.
\end{aligned}
\end{equation}
Since $2^{{k}n}q_n^{k}$ only grows like a power of
$n$ while the number of choices for $\bsd^{t_1}$ is bounded
by $2^{k}$
and $n_{\bsd^1}=n2^{-k} +o(n)$ by the bound
\eqref{const-lan},
we conclude
that the contribution of the regions where
at least one of the $|v_s|$'s
is larger than $\mu_1$ is exponentially small in $n$.

Consider now the region where all $|v_s|$'s are bounded
by $\mu_1$.  In this region, we again
would like to approximate the $\sinc$ factors in
\eqref{onetwo} by one.
To this end, we first note
that
\begin{equation}
\label{v-f}
\sum_{j=1}^k|f_j|=\max_{s\leq n} |v_s|.
\end{equation}
Indeed, by the
triangle inequality, we clearly have that
$\max_s|v_s|\leq \sum_j|f_j|$.  To prove the opposite
inequality, we use that
$n_{\bsd} = \frac{n}{2^{k}}(1 + o(1)) > 0$ for every $\bsd \in
\{-1,+1\}^n$, implying that there exists a $v_{s_0}$ that is
evaluated as $\sum_{j=1}^{k} \abs{f_j} $.
In the region where all $|v_s|$'s are bounded
by $\mu_1$, we therefore have that all $f_j$'s are
bounded by $\mu_1$,
so that
$ \sinc(q_n f_j)=1+O (q_n^2)$.  By the fact that
$t_n\sqrt n=\alpha\sqrt n +O(q_n)$,
we furthermore
have that $\cos(2\pi f_j t_n \sqrt{n})=\cos(2\pi f_j\alpha
\sqrt{n})+O(q_n)$.
Replacing the $\sinc$ factors by $1$, and the product of
$\cos(2\pi f_j t_n \sqrt{n})$ by  $\cos(2\pi f_j \alpha \sqrt{n})$,
we therefore obtain an error which can be bounded by
$2^{n{k}}q_n^k O(q_n)$, an error which again goes to zero exponentially
in $n$.

We thus have show that, up to an error which is exponentially small
in $n$, the left hand side of \eqref{Ik-asympt} is equal to
\begin{equation}
\label{onetwo-bd}
2^{n{k}}q_n^{k}\iiint_{-\mu_1}^{\mu_1} \;  \prod_{s=1}^n
\rhat(v_s) \prod_{j=1}^{k}\cos(2\pi f_j \alpha \sqrt{n})
df_j.
\end{equation}
Next we show that  we can further restrict the range
of integration to $|v_s|\leq \omega_n/\sqrt n$ for all $s$.
To this end,
let us consider the integral where $\omega_n\leq |v_{s_1}|\leq \mu_1$,
while $v_s$ can be an arbitrary number in $[-\mu_1,\mu_1]$ for
all other $s$.  We will then have to bound the integral
\begin{equation}
\notag
\tilde I_3 =  \iiint_{-\mu_1}^{\mu_1}
\int_{\omega_n/\sqrt n\leq \abs{v_{t_1}}\leq
\mu_1} \; J_{k} \prod_{s=1}^n \rhat(v_s)
\prod_{j=1}^{k} \cos(2\pi f_j \alpha \sqrt{n})  dv_{t_j}.
\end{equation}
Using the fact that $\rhat(v)\leq e^{-c_2 v^2}$
for $|v|\leq \mu_1$, this can be easily accomplished, leading to
the bound
\begin{equation}
\label{onetwo-3}
\begin{aligned}
|\tilde I_3|
&\leq \iiint_{-\infty}^{\infty}
\prod_{j=2}^{k} e^{-n_{\bsd^j}c_2 v_{t_j}^2} dv_{t_j}
\times \int_{\abs{v_{t_1}}
>\omega_n/\sqrt n} e^{-n_{\bsd^1}c_2 v_{t_1}^2} dv_{t_1}
\\
&=O(n^{-{k}/2}e^{-c_2\omega_n^2(n_{\bsd^1}/ n)}).
\end{aligned}
\end{equation}
Using the facts that $\omega_n\to\infty$,
$n_{\bsd^1}/ n=2^{-{k}}+o(1)$ and
$2^{{k}n}q_n^{k}=O(n^{{k}/2})$,
this implies that over all $2^{nu}$
sequences of configurations $\bss^{(1)},\dots,\bss^{(u)}\in\{-1,+1\}^n$,
the contribution of the regions where
$|v_{s_1}|$ is larger than $\omega_n/\sqrt n$ is negligible.
However, since there are at most $2^k$ different possibilities
for $|v_s|$, we  see that the contribution of the regions where
any one of the $|v_s|$'s
is larger than $\omega_n/\sqrt n$ is negligible.

For $|v_s|\leq\omega_n/\sqrt n$, we approximate
$\rhat(v_s)$ by $\rhat(v_s)=  \exp(-2\pi^2 v_s^2+o(1/n))$.
Using the shorthand $n_\bsd$ for the quantity
$n_\bsd(\bss^{(1)},\dots,\bss^{({k})})$, and defining
$v_\bsd$ as $\sum_{j=1}^k \delta_j f_j$, we then rewrite
$$
\prod_{s=1}^n\rhat(v_s)
=\prod_{\bsd}
\rhat(v_\bsd)^{n_\bsd}
=\exp\Bigl(
-2\pi^2\sum_{\bsd} n_{\bsd}v_{\bsd}^2+o(1)
\Bigr).
$$
We would like to approximate the sum in the
exponent by
$f^2=\sum_{j=1}^{k} f_j^2$.
To this end, we first note that
\[
\sum_{\bsd \in \{-1,+1\}^k} \sum_{j_1,j_2} \delta_{j_1}
f_{j_1} \delta_{j_2} f_{j_2} = \sum_{\bsd \in
\{-1,+1\}^k} \sum_{j} f_{j}^2 =2^k\sum_{j} f_{j}^2
=2^k f^2.
\]
If $n_\bsd$ was \emph{equal} to $2^{-k}n$ for all $\bsd$, the sum
in the exponent would therefore be equal to $f^2$, but for general
$\bsd$ we get the bound
\begin{equation}
\label{sum-nd-f}
\begin{aligned}
\biggl|\sum_{\bsd} n_{\bsd}v_{\bsd}^2
-nf^2\biggr|
&=
\biggl|\sum_{\bsd} (n_{\bsd}-2^{-k}n)v_{\bsd}^2
\biggr|
\leq
\biggl(\max_\bsd \bigl|n_{\bsd}-2^{-k}n\bigr|\biggr)
\sum_{\bsd} v_{\bsd}^2 .
\end{aligned}
\end{equation}
Using the condition \eqref{const-lan}, and the fact
that $|v_\bsd|\leq \omega_n/\sqrt n$, we bound the right
hand side by
$\lambda_n2^k\omega_n^2/\sqrt n=o(1)$.
We thus have shown that
for $|v_s|\leq\omega_n/\sqrt n$,
\[
\begin{aligned}
\prod_{s=1}^n\rhat(v_s)
=\bigl(1+o(1)\bigr)
\exp\Bigl(
-2\pi^2nf^2\Bigr).
\end{aligned}
\]

Combining the bounds proven so far, we conclude that,
up to an error which is negligible as $n\to\infty$,
the expression in \eqref{onetwo} is  equal to
\begin{equation}
\label{Ik-approx}
 2^{n{k}}q_n^{k}\iiint \;
 \prod_{j=1}^{k} \exp(
-2\pi^2nf_j^2)\cos(2\pi f_j \alpha \sqrt{n})
df_j,
\end{equation}
where the integral goes the region where
$|v_s|\leq \omega_n/\sqrt n$ for all $s$.  Since,
by an argument very similar to the argument leading to
\eqref{onetwo-1} and
\eqref{onetwo-3}, the
integral of $ \prod_{j=1}^{k} \exp(
-2\pi^2nf_j^2)$ over a region in which
$|v_s|> \omega_n/\sqrt n$ for at least one $s$ is negligible,
we therefore have  shown that
\begin{equation}
\begin{aligned}
2^{n{k}}\mathbb{E}&[I_{k}(\bss^{(1)},\dots, \bss^{({k})})]
\\
&=
 2^{n{k}}q_n^{k}\iiint_{-\infty}^\infty \;
 \prod_{j=1}^{k} \exp(
-2\pi^2nf_j^2)\cos(2\pi f_j t_n \sqrt{n})
df_j + o(1)
\\
&=\gamma^{k} +o(1),
\end{aligned}
\end{equation}
as desired.
\end{proof}

As we will see below, the number sequences
of configurations $\bss^{(1)},\dots, \bss^{({k})}$
that are linearly independent and satisfy the bound \eqref{const-lan}
is $2^{nk} (1+o(1))$.  Restricting the sum in
\eqref{Zm} to these configurations and using
 Lemma~\ref{lem:Zkk-asymp} to estimate the expectation
 of $I_{k}(\bss^{(1)},\dots, \bss^{({k})})]$,
we get a contribution to the
 $k^{\text{th}}$ factorial moment that is asymptotically
 equal to $\gamma^k$.  To prove Theorem~\ref{main-thm},
 we have to bound the contribution of the remaining terms.
 To this end, we first establish an upper bound on the
 expectation of
 $I_{k}(\bss^{(1)},\dots, \bss^{({k})})]$ that does not
 rely on the condition \eqref{const-lan}.
 To formulate this bound, we introduce the following notation.

\begin{definition}
\label{def:n0-rank} Let $n_0$ be such that $ 1/n_0 + 1/(1+\epsilon) =
1$ where $\epsilon$ is the constant  from assumption \eqref{dist}.
We say that the configurations $\bss^{(1)},...,\bss^{(u)}$
has $n_0$-rank $u_0$ if the maximum number of
linearly independent column
vectors $\bsd\in\{-1,+1\}^u$ such that
\begin{equation}
\label{condA} n_{\bsd}(\bss^{(1)},...,\bss^{(u)}) \geq n_0
\end{equation}
is equal to $u_0$.
\end{definition}

\begin{lemma}
\label{lem:Iu-bd} Given  a positive integer $u$,
there exists a constant
$C_u$  such that for
all
sets of linearly independent row vectors
 $\bss^{(1)},...,\bss^{(u)}\in\{-1,+1\}^n$
that have $n_0$-rank $u_0$, we have
\begin{equation}
\label{Iubd}
 \left| \mathbb{E}[I_u(\bss^{(1)},\dots, \bss^{(u)})]
\right| \leq C_uq_n^{u_0+(u-u_0)/n_0}.
\end{equation}
\end{lemma}

\begin{proof}
Let $A_\bsd\subset\{1,\dots,n\}$ be the set of indices
$i$ such that the column vector $(\sigma_i^{(1)},
\dots, \sigma_i^{(u)})$ is equal to $\bsd$,
and let $\tilde Y_\bsd$ be the random variable
\begin{equation}
\tilde Y_\bsd = \sum_{i\in A_\bsd} Y_i.
\end{equation}
Recalling the definition \eqref{Im},
we then rewrite
$
I_u(\bss^{(1)},\dots, \bss^{(u)})
$
as
\begin{align}
\label{Iu}
&I_u(\bss^{(1)},\dots, \bss^{(u)})=  \\
& \quad =
2^{-u}\sum_{\boldsymbol\tau\in \{-1,+1\}^u}
  \prod_{j=1}^u
  \rect(\frac{\sum_{\bsd\in\Delta}
  \delta_j \tilde Y_\bsd -\tau_jt_n\sqrt n}{q_n})
   \notag
\end{align}
where $\Delta$ is the set of vectors $\bsd\in\{-1,+1\}$
such that $n_\bsd\geq 1$.

Choose $u$ linearly independent vectors
$\bsd^{(1)},\dots,\bsd^{(u)}\in\Delta$
such that the vectors
$\bsd^{(1)},\dots,\bsd^{(u_0)}$
satisfy the condition \eqref{condA}.
Let $\Delta_0=\{\bsd^{(1)},\dots,\bsd^{(u_0)}\}$,
and let $\Delta_u=\{\bsd^{(1)},\dots,\bsd^{(u)}\}$.
Denoting the $k$-fold convolution of $\rho$
with itself by $\rho_k$, we then write the expectation
of a typical term on the right hand side of
\eqref{Iu} as
\begin{equation}
\label{EIu}
\mathbb E
\Bigl[
\prod_{j=1}^u
  \rect(\frac{\sum_{\bsd\in\Delta}
  \delta_j \tilde Y_\bsd -\tau_jt_n\sqrt n}{q_n})
\Bigr]
=
\iiint K_u(y_{\Delta\setminus\Delta_0})
\prod_{\bsd\in \Delta\setminus\Delta_u}
\rho_{n_\bsd}( y_\bsd) dy_\bsd
\end{equation}
where
$y_{\Delta\setminus\Delta_0}$ is a shorthand for the collection
of variables $y_\bsd$, $\bsd\in\Delta\setminus\Delta_0$, and
$K_u(y_{\Delta\setminus\Delta_0})$ is the integral
\begin{equation}
\begin{aligned}
K_u(y_{\Delta\setminus\Delta_0})&=
\iiint
  \prod_{j=1}^u
  \rect(\frac{\sum_{\bsd\in\Delta} \delta_j
  y_\bsd -\eta_j t_n\sqrt{n}}{q_n})
 \prod_{\bsd\in \Delta_u}
\rho_{n_\bsd}(y_\bsd) dy_\bsd .
   \notag
\end{aligned}
\end{equation}
Combining the relations \eqref{Iu} and \eqref{EIu}
and observing that
\[
\iiint
\prod_{\bsd\in \Delta\setminus\Delta_u}
\rho_{n_\bsd}( y_\bsd) dy_\bsd=1
\]
by the fact that $n_\bsd\geq 1$ for all $\bsd\in\Delta$,
we clearly have that
\begin{equation}
 \left| \mathbb{E}[I_u(\bss^{(1)},\dots, \bss^{(u)})]
\right| \leq
\sup_{y_{\Delta\setminus\Delta_0}\in\R^{|\Delta\setminus\Delta_0|}}
K_u(y_{\Delta\setminus\Delta_0}).
\end{equation}
It is therefore enough to bound
$K_u(y_{\Delta\setminus\Delta_0})$ uniformly in
$y_{\Delta\setminus\Delta_0}$.

Let $\alpha_j=\tau_jt_n\sqrt n
-\sum_{\bsd\in\Delta\setminus\Delta_u}\delta_jy_\bsd$.
Noting that $\alpha_j$ does not depend on the variables
which are integrated over in $K_u$, we
then rewrite $K_u$ as
\begin{equation}
\begin{aligned}
K_u(y_{\Delta\setminus\Delta_0})&=
\iiint
  \prod_{j=1}^u
  \rect(\frac{\sum_{\bsd\in\Delta_u} \delta_j
  y_\bsd -\alpha_j}{q_n})\prod_{\bsd\in \Delta_u}
\rho_{n_\bsd}(y_\bsd) dy_\bsd
   \notag
\end{aligned}
\end{equation}
Let $\tilde M$ be the matrix with matrix elements
$\tilde M_{ji}=\delta^{(i)}_j$.  The product of the $\rect$-functions
in the above integral then ensures that
\begin{equation}
\label{alpha-bd}
\max_{j=1,\dots,u}
\Bigl|\sum_{i=1}^u M_{ji} y_{\bsd^{(i)}}-\alpha_j\Bigr|
\leq \frac 12 q_n  .
\end{equation}
Since the vectors in $\Delta_u$ are linearly independent,
the matrix $\tilde M$ is invertible.  Let
$\beta_i=\sum_{j=1}^u(\tilde M^{-1})_{ij}\alpha_j$, and let
$\|\tilde M^{-1}\|$ be the norm of $\tilde M^{-1}$ as an operator
from $\ell_\infty$ to $\ell_\infty$.
The bound \eqref{alpha-bd} then implies that
\begin{equation}
\label{beta-bd}
\max_{i=1,\dots,u}
\Bigl|y_{\bsd^{(i)}}-\beta_i\Bigr|
\leq \frac 12 \tilde q_n,
\end{equation}
where $\tilde q_n=\|\tilde M^{-1}\| q_n$.
As a consequence, the integral $K_u$ is bounded by
\begin{equation}
\begin{aligned}
K_u(y_{\Delta\setminus\Delta_0})&\leq
\iiint
  \prod_{i=1}^u
  \rect(\frac{
  y_{\bsd^{(i)}}-\beta_i}{\tilde q_n})\prod_{\bsd\in \Delta_u}
\rho_{n_\bsd}(y_\bsd) dy_\bsd
  \\
  &=
  \prod_{i=1}^u
  \int  \rect(\frac{  y-\beta_i}{\tilde q_n})
\rho_{n_i}(y) dy
\\
&=
 \tilde q_n^u \prod_{i=1}^u
 \int \rhat^{n_i}(f) \sinc(q_n f) e^{2\pi i \beta_i f}df,
\end{aligned}
\end{equation}
where we used the shorthand $n_i=n_\bsd^{(i)}$.  For $i=1,\dots,u_0$,
we use that $n_i\geq n_0$ to bound the integral on the right by
\begin{equation}
\int \rhat^{n_i}(f) \sinc(q_n f) e^{2\pi i \beta_i f}df
\leq \int |\rhat^{n_0}(f)|df
=C_0,
\end{equation}
while for $i=u_0+1,\dots u$, we use $n_i\geq 1$
and H\"older's inequality to obtain the bound
\begin{equation}
\begin{aligned}
\int \rhat^{n_i}(f) &\sinc(q_n f) e^{2\pi i \beta_i f}df
\leq
\int |\rhat(f)\sinc (\tilde q_n f)|df
\\
&\leq \Bigl[\int |\rhat^{n_0}(f)|df\Bigr]^{1/n_0}
\Bigl[\int |\sinc (\tilde q_n f)|^{1+\epsilon}df\Bigr]^{1/(1+\epsilon)}
\\
&=\tilde C_0 \tilde q_n^{-1/(1+\epsilon)}.
\end{aligned}
\end{equation}
Here
\begin{equation}
\tilde C_0 =
\Bigl[\int |\rhat^{n_0}(f)|df\Bigr]^{1/n_0}
\Bigl[\int |\sinc (f)|^{1+\epsilon}df\Bigr]^{1/(1+\epsilon)}<\infty
\end{equation}
is independent of $u$, $u_0$ and $n$.  Observing that
$1-1/(1+\epsilon)=1/n_0$, we thus get the bound
\begin{equation}
K_u(y_{\Delta\setminus\Delta_0})
\leq (\max C_0,\tilde C_0)^u \tilde q_n^{u_0+(u-u_0)/n_0}.
\end{equation}
Since
there is only a finite number of choices for
a set $\Delta$ of $u$ linearly independent vectors
in $\{-1,+1\}^u$, the ratio $\tilde q_n/q_n=\|\tilde M^{-1}\|$
is bounded by a constant that depends only on $u$, implying the
existence of a constant $C_u$ such that
\begin{equation}
K_u(y_{\Delta\setminus\Delta_0})
\leq C_u q_n^{u_0+(u-u_0)/n_0}.
\end{equation}
Combined with \eqref{Iu} and \eqref{EIu}, this proves the lemma.
\end{proof}

In order to bound the contribution  in equation (\ref{Zm})
coming from the terms where the vectors $\bsd^{(1)},...,\bsd^{(k)}$
have rank $u < k$ or do not satisfy  condition
\eqref{const-lan}, we need the following lemma, whose main statements
were already proven in
\cite{borgs:chayes:pittel:01}.

\begin{lemma}
\label{lem:comb-bds}
\noindent \begin{enumerate}
\item
Given $u \leq k$ linearly independent row vectors
$\bss^{(1)},\dots,\bss^{(u)}$, there are at most $2^{u(k-u)}$ ways
to choose $\bss^{(u+1)},\dots,\bss^{(k)}$ such that
the matrix
$M$ formed by the row vectors $\bss^{(1)},\dots,\bss^{(k)}$ has rank
$u$.

\item Given $u$ and $n_0$, there are constants
$c_3=c_3(u,n_0)$ and $C_3=C_3(u,n_0)$ such that there are at most
$C_3n^{c_3}2^{nu_0}$
ways to choose $u$ linearly independent
configurations $\bss^{(1)},...,\bss^{(u)}$
that have $n_0$-rank $u_0$.

\item  Let $u<\infty$, let $c_4=c_4(u)=2^{u+1}$, and let
$\lambda_n$ be a sequence of positive number such that
$\lambda_n/\sqrt n\to 0$ as $n\to\infty$.  Then
the number of configurations
$\bss^{(1)},...,\bss^{(u)}$ that violate condition
\eqref{const-lan}
is bounded by
$c_42^{nu}e^{-\frac 12\lambda_n^2}$.

\item
Let  $\bss^{(1)},...,\bss^{(k)}$ be distinct spin
configurations, assume that rank  $M<k$,
and let $\bss^{(1)},...,\bss^{(u)}$ be
linearly independent.
Then
$n_\bsd(\bss^{(1)},...,\bss^{(u)})=0$ for at least one
$\bsd\in\{-1,+1\}^u$, implying in particular that, for $n$
sufficiently large, $\bss^{(1)},...,\bss^{(u)}$ violate condition
\eqref{const-lan}.

\item Given $u$, let
$\bss^{(1)},...,\bss^{(u)}\in\{-1,+1\}^n$ be
an arbitrary set of row vectors
satisfying \eqref{const-lan}.  Then
\begin{equation}
\label{olap}
q(\bss^{(a)},\bss^{(b)})
\leq
2^u\frac{\lambda_n}{\sqrt n}
\end{equation}
whenever $a\neq b$.  For
$n$ sufficiently large,  condition \eqref{const-lan}
therefore implies that
$\bss^{(1)},...,\bss^{(u)}$ are linearly independent.
\end{enumerate}
\end{lemma}

\begin{proof}
Except for the second statement, the
lemma mainly summarizes
the relevant results from Section 6 of
\cite{borgs:chayes:pittel:01}.  More explicitly:
statement (1) is proved
in the paragraph following (6.10), and for
$\lambda=\log n$, statement (3), is
proved in the paragraphs around
(6.12),
statement (4) is proved in the
paragraph around the second and third unnumbered equation after
(6.12),
and statement (5) is equivalent to the bound (6.14).

It is not hard to see that the arguments in
Section 6 of
\cite{borgs:chayes:pittel:01} can be generalized to
arbitrary sequences
$\lambda_n$ of positive number, as long as
 $\lambda_n/\sqrt n\to 0$ as $n\to\infty$.
Indeed, starting with statement (3), let us consider $n$
independent trials with $2^u$ equally likely outcomes, and use
Chebychev's inequality to bound the probability that
$|n_{\bsd}(\bss^{(1)},...,\bss^{(u)}) - {n}{2^{-u}}|
\geq\sqrt{n}\lambda_n$ by $2e^{-\frac 12\lambda^2_n}$. Combined with the
union bound for the $2^u$ different random variables
$n_{\bsd}(\bss^{(1)},...,\bss^{(u)})$, $\bsd\in\{-1,+1\}^u$, this
gives statement (3).
The first part of (4) does not involve the value of
$\lambda_n$, and the second follows from the first whenever
$\lambda n/\sqrt n\to 0$ as $n\to\infty$.  To prove
the bound \eqref{olap} in statement (5), we rewrite
 the overlap $q(\bss^{(a)},\bss^{(b)})$ as
\begin{equation}
\begin{aligned}
&q(\bss^{(a)},\bss^{(b)})
=
\\
&\quad=\frac 1n
\biggl(\sum_{\bes\bsd\in\{-1,+1\}^u \\ \delta_a=\delta_b\es}
n_{\bsd}(\bss^{(1)},...,\bss^{(u)})
-
\sum_{\bes\bsd\in\{-1,+1\}^u \\ \delta_a\neq\delta_b\es}
n_{\bsd}(\bss^{(1)},...,\bss^{(u)})
\biggr).
\end{aligned}
\end{equation}
Noting that each sum contains $2^{u-1}$ terms,
we see that the bound \eqref{const-lan} implies the bound
\eqref{olap}.  Finally, the last statement of (5) is a
direct consequence of \eqref{olap} and the fact that
$\lambda_n/\sqrt n\to 0$ as $n\to\infty$.

We are left with the proof of (2).  To this end, let us
consider the matrix $\tilde M_u$ obtained from $M_u$
by omitting all columns $\sigma_i^{(1)},\dots,\sigma_i^{(u)}$
that are equal to a vector $\bsd\in\{-1,+1\}^u$ with
$n_{\bsd}(\bss^{(1)},...,\bss^{(u)})<n_0$.  Note that
the number of columns $n'$ of
$\tilde M_u$ is at least $n-2^un_0$ and at most
$n$.  Fixing $n'$, for the moment, and noting that
the rank of $\tilde M_u$ is $u_0$, we now use statement (1)
to conclude that there are at most
\begin{equation}
\binom{u}{u_0}2^{n'u_0}2^{u_0(u-u_0)}
\leq 2^{u+u^2/2} 2^{nu_0}
\end{equation}
ways to choose $\tilde M_u$.
Given $\tilde M_u$ we need to insert $n-n'$ columns in
$\{-1,+1\}^u$ to obtain the matrix $M_u$.  Including the
number of choices for the positions of these $n-n'$
columns, this gives an extra factor of
\begin{equation}
\binom{n}{ n-n'}2^{(n-n')u}
\leq \frac 1{(n-n')!}n^{n-n'}2^{n-n'}
\leq \frac 1{(n-n')!}n^{n_02^u}2^{n_02^u}
\end{equation}
Combining the two factors and summing over $n'\in\{n-n_02^u,\dots,n\}$,
we get a bound of the form $C_3 n^{c_3} 2^{u_0n}$ where
$C_3$ and $c_3$ depend only on $u$ and $n_0$.
\end{proof}

Having Lemmas~\ref{lem:int-exch-u}, \ref{lem:Zkk-asymp}, \ref{lem:Iu-bd} and
\ref{lem:comb-bds} in hand, we are now ready to
prove Theorem~\ref{main-thm}.

\subsubsection{Proof of Theorem~\ref{main-thm}}

We start with the case $m=1$, i.e., the one-dimensional factorial
moment $\mathbb E[( Z_n(a_n,b_n))_k]$.
As in Lemma~\ref{lem:Zkk-asymp}, we choose $\lambda_n=\log n$.
Consider the sum over all sequences
$\bss^{(1)},\dots, \bss^{({k})}\in\{-1,+1\}^n$ that
satisfy the bound \eqref{const-lan}.
By Lemma~\ref{lem:comb-bds} (3), this sum contains
$2^{nk} (1+o(1))$ terms, and by
Lemma~\ref{lem:comb-bds} (5), the matrix formed by the
row vectors $\bss^{(1)},\dots, \bss^{({k})}$ has rank $k$
if $n$ is large enough.  With the help of
Lemma~\ref{lem:Zkk-asymp}, we conclude that
the sum over all these terms gives a contribution to the
$k^{\text{th}}$ factorial moment which is
equal to $\gamma^k+o(1)$.

To prove Theorem~\ref{main-thm},
we have to bound the contribution of the remaining terms.
To this end, we
group the remaining terms in the
sum \eqref{Zm} into  four
classes:

\begin{enumerate}

\item Sequences of distinct configurations
$\bss^{(1)},\dots, \bss^{({k})}$ of rank
$k$ and $n_0$-rank $u_0<k$
that violate the condition \eqref{const-lan};

\item Sequences of distinct configurations
$\bss^{(1)},\dots, \bss^{({k})}$ of rank $k$ and $n_0$-rank $k$
that violate the condition \eqref{const-lan};

\item
Sequences of distinct configurations
$\bss^{(1)},\dots, \bss^{({k})}$ of rank $u<k$
such that there is a subsequence of
linearly independent configurations
$\tilde \bss^{(1)},\dots, \tilde\bss^{({u})}$
of $n_0$-rank $u_0<u$;

\item
Sequences of distinct configurations
$\bss^{(1)},\dots, \bss^{({k})}$ of rank $u<k$
such that  all  subsequences of
linearly independent configurations
$\tilde \bss^{(1)},\dots, \tilde\bss^{({u})}$
have
of $n_0$-rank $u_0=u$;

\end{enumerate}

By Lemma~\ref{lem:comb-bds} (4), the configurations
$\tilde \bss^{(1)},\dots, \tilde\bss^{({u})}$
in class (4) must violate  condition \eqref{const-lan}.
Relaxing the constraint that the configurations in
class (1) violate  condition \eqref{const-lan}, it is
therefore enough to bound the
following two error terms:
\begin{itemize}
\item
the sum $R_{n,k}^{<}$ of all sequences of configurations
$\bss^{(1)},\dots, \bss^{({k})}$ of rank $u\leq k$
containing a subsequence
of linearly independent configurations
$\tilde \bss^{(1)},\dots, \tilde\bss^{({u})}$
of $n_0$-rank $u_0<u$, and
\item the sum
$R_{n,k}^{=}$ of all sequences of configurations
$\bss^{(1)},\dots, \bss^{({k})}$ of rank $u\leq k$
such that all subsequence
of linearly independent configurations
$\tilde \bss^{(1)},\dots, \tilde\bss^{({u})}$
obey  condition \eqref{const-lan} and have
$n_0$-rank $u_0=u$.
\end{itemize}

Before bounding these two error terms, we  note that
\begin{equation}
I_k(\bss^{(1)},\dots, \bss^{(k)})
=\prod_{i=1}^k I(\bss^{(i)})
\leq
\prod_{i=1}^u I(\tilde\bss^{(i)})
=I_u(\tilde\bss^{(1)},\dots, \tilde\bss^{(u)}),
\end{equation}
implying that
\begin{equation}
\label{Zk2}
\mathbb E[I_k(\bss^{(1)},\dots, \bss^{(k)})]
\leq
\mathbb E[I_u(\tilde\bss^{(1)},\dots, \tilde\bss^{(u)})]
\end{equation}
whenever $\tilde \bss^{(1)},\dots, \tilde\bss^{({u})}$ is
a subsequence of $\bss^{(1)},\dots, \bss^{(k)}$.

In order to bound $R_{n,k}^{<}$, we now use
\eqref{Zk2} and Lemma~\ref{lem:Iu-bd} to bound
the expectation of
$I_k(\bss^{(1)},\dots, \bss^{(k)})$
by $C_uq_n^{u_0+(u-u_0)/n_0}\leq c_5 q_n^{u_0} q_n^{1/n_0}$,
where $c_5=\max_{u\leq k} C_u$.
Using Lemma~\ref{lem:comb-bds} (2) to bound the number
of sequences
$\tilde \bss^{(1)},\dots, \tilde\bss^{({u})}$
of $n_0$-rank $u_0$ by $C_6 n^{c_6}2^{nu_0}$,
where $C_6=\max_{u\leq k}C_3(u,n_0)$ and $c_6=\max_{u\leq k}c_3(u,n_0)$,
and Lemma~\ref{lem:comb-bds} (1) to bound the number of ways
$\bss^{(1)},\dots, \bss^{(k)}$ can be obtained from
$\tilde \bss^{(1)},\dots, \tilde\bss^{({u})}$, we therefore obtain
the following upper bound
\begin{equation}
R_{n,k}^{<}
\leq C_6c_5n^{c_6}\sum_{\bes u_0,u:\\ u_0<u\leq k\es}
\binom{k}{u}2^{u(k-u)}
 (2^nq_n)^{u_0} q_n^{1/n_0}.
\end{equation}
Since $2^n q_n=O(\sqrt n)$, we get that
\begin{equation}
\label{R<bd}
R_{n,k}^{<}=O(n^{c_7}q_n^{1/n_0})
\end{equation}
where the constant implicit in the $O$-symbol depends on
$k$, $\alpha$ and $\gamma$, and where $c_7=c_6+k/2$.  Since $q_n$ falls
exponentially with $n$, this proves that
$R_{n,k}^{<}=o(1)$.

The error term $R_{n,k}^{=}$ can be bounded in a similar way.
We again use \eqref{Zk2} and Lemma~\ref{lem:Iu-bd} to bound
the expectation of
$I_k(\bss^{(1)},\dots, \bss^{(k)})$,
but now we use part (3) of
Lemma~\ref{lem:comb-bds} to bound the number of
sequences
$\tilde \bss^{(1)},\dots, \tilde\bss^{({u})}$.
Using again
Lemma~\ref{lem:comb-bds} (1) to bound the number of ways
$\bss^{(1)},\dots, \bss^{(k)}$ can be obtained from
$\tilde \bss^{(1)},\dots, \tilde\bss^{({u})}$, we now obtain
the  upper bound
\begin{equation}
R_{n,k}^{=}
\leq c_5c_8\sum_{\bes u_0,u:\\ u_0<u\leq k\es}
\binom{k}{u}2^{u(k-u)}
e^{-\lambda_n^2/2}
 (2^nq_n)^{u}
\end{equation}
where $c_8=\max{u\leq k}c_4(u)=2^{k+1}$.  Using again
that $2^n q_n=O(\sqrt n)$, we conclude that
\begin{equation}
\label{R=bd}
R_{n,k}^{=}=O(n^{k/2}e^{-\lambda_n^2/2}).
\end{equation}
Since $e^{-\lambda_n^2/2}=e^{-\log^2n/2}$ decays faster than any
power of $n$, the right hand side goes to zero as $n\to\infty$,
as desired.

This completes the proof that $\mathbb E[( Z_n(a_n,b_n))_k]\to\gamma^k$
as $n\to\infty$. To prove the convergence of the higher-dimensional
factorial moments, we need to generalize Lemmas~\ref{lem:int-exch-u},
\ref{lem:Zkk-asymp}
and \ref{lem:Iu-bd}.
But, except for notational inconveniences, this causes no problems.
Indeed,
comparing the representations \eqref{Im} and \eqref{Im-multi},
we see that the only difference is the appearance of several
distinct intervals $[a_n^{\ell(j)},b_n^{\ell(j)}]$
for the energy of the configuration $\bss^{(j)}$, instead of
the same interval $[a_n,b_n]$ for all of them.

As a consequence, the statement of
Lemma~\ref{lem:int-exch-u}
has to be modified, with the right hand side of \eqref{int3}
replaced by
\begin{equation}
\label{int3-multi}
\prod_{\ell=1}^k q_{n,\ell}^{k_\ell}
\iiint_{-\infty}^{\infty} \;  \prod_{s=1}^n \rhat(v_s)
\prod_{j=1}^{k} \sinc(f_j q_{n,\ell(j)})
\cos(2\pi f_j t_n^{\ell(j)} \sqrt{n})df_j.
\end{equation}
But the proof remains unchanged, since
it never used that $q_{n,\ell(j)}$ or $t_n^{\ell(j)}$ is
constant.

In a similar way, the proof of
Lemma~\ref{lem:Zkk-asymp} needs only notational changes:
the arguments leading
to \eqref{onetwo-bd} now give a prefactor
$2^{nk}\prod_j q_{\ell(j)}$ instead of
$2^{nk}q_n^k$, but the integral multiplying this prefactor
(and therefore the rest of the proof) remains
unchanged, proving that under the conditions of
Lemma~\ref{lem:Zkk-asymp},
\eq
2^{n{k}}\mathbb{E}[I_{k}(\bss^{(1)},\dots, \bss^{({k})})]
=\prod_j \gamma_{\ell(j)} +o(1).
\en

Turning finally to the proof of
Lemma~\ref{lem:Iu-bd}, we note that its proof goes through if we
replace $\tilde q_n$
by $\|\tilde M^{-1}\|\max_{\ell=1,\dots,m} q_{n,\ell}$.
As a consequence, the bound \eqref{Iubd}  has to be modified to
\begin{equation}
\label{Iubd2}
 \left| \mathbb{E}[I_u(\bss^{(1)},\dots, \bss^{(u)})]
\right| \leq C_u(\max_{\ell=1,\dots,m}q_{n,\ell})^{u_0+(u-u_0)/n_0},
\end{equation}
which does not change the $n$-dependence of the bound.

Using these generalizations of
Lemmas~\ref{lem:int-exch-u},
\ref{lem:Zkk-asymp}
and \ref{lem:Iu-bd}, it is easy to see that the bounds
\eqref{R<bd} and \eqref{R=bd} remain unchanged,
except for the fact that
the implicit constants in the $O$-symbols
now depend on $\max_\ell\gamma_\ell$ instead of
$\gamma$.  This completes the convergence proof for
the  multi-dimensional
factorial moments, and hence the proof of
Theorem~\ref{main-thm}.

\begin{remark}
Throughout this section, we have assumed that
$\alpha$ is bounded.  For convenience in our
companion paper \cite{BCMN-2}, we note that
the above estimates on $R_{n,k}^{<}$ and $R_{n,k}^{=}$
can be easily be generalized to growing $\alpha$ if we
choose $\lambda_n$ appropriately. Indeed, making the
$\alpha$-dependence of our bounds explicit, we obtain
\begin{equation}
R_{n,k}^{<}=O(n^{c_7}e^{k\alpha^2/2}q_n^{1/n_0})
\end{equation}
and
\begin{equation}
R_{n,k}^{=}=O(n^{k/2}e^{k\alpha^2/2}e^{-\lambda_n^2/2}),
\end{equation}
as long as $\lambda_n=o(\sqrt n)$.  For $\alpha=o(\sqrt n)$,
$q_n$ decays exponentially in $n$, and $R_{n,k}^{<}=o(1)$.
Choosing $\lambda_n$ in such a way that $\alpha=o(\lambda_n)$,
$\lambda_n=o(\sqrt n)$ and
$e^{-\lambda_n^2/2}$ decays faster than any power of $n$,
we also have $R_{n,k}^{=}=o(1)$.

In order to prove Theorem~\ref{main-thm}
for growing $\alpha$, we therefore only
need to generalize the statements of Lemmas~\ref{lem:Z-asymp}
and \ref{lem:Zkk-asymp}.  For $\alpha=o(n^{1/4})$,
this will be done in \cite{BCMN-2}.
\end{remark}

\subsection{Overlap Estimates}

To complete the proof of the Theorem \ref{mainth}, we need to show
that the rescaled overlaps converge to a standard normal.
Defining $R(\beta)$ to be the tail of the standard Gaussian,
\[ R(\beta) = \int_{\beta}^{\infty} \frac{1}{\sqrt{2\pi}}
e^{-\frac{x^2}{2}} dx.
\]
we therefore have to show that for any $\beta\in R$ and
any $j>i>0$, we have
\begin{equation}
\label{PQ-lim}
\PR(Q_{ij}\geq \beta)\to R(\beta)
\end{equation}
as $n\to\infty$.

Let $E_{r_n+i}$ and $E_{r_n+j}$ be the $i^{\text{th}}$ and
$j^{\text{th}}$ energy above $\alpha$, respectively,
and let $\lambda_0>0$.
Having established the convergence \eqref{rem-general} of the
rescaled energies, we note that
the probability that
both $E_{r_n+i}$ and $E_{r_n+j}$ fall into the interval
$[\alpha,\alpha+\lambda_0\xi_n]$ can be made arbitrary close to
one by choosing  $\lambda_0$ and  $n$ large enough.
Consider further a
discretization scale $\eta$ such that $\lambda_0/\eta$ is an
integer. If both $E_{r_n+i}$ and $E_{r_n+j}$ fall into the interval
$[\alpha,\alpha+\lambda_0\xi_n]$, each of them must fall into
one of the $\lambda_0/\eta$ intervals $[\alpha,\alpha+\eta\xi_n]$,
$[\alpha+\eta\xi_n,\alpha+2\eta\xi_n]$, $\dots$,
$[\alpha+(\gamma_0-\eta)\xi_n,\alpha+\lambda_0\xi_n]$.
By choosing $\eta$ sufficiently small and $n$ sufficiently large,
the probability that both fall into the
same interval, or that one of the other energies between
$\alpha$ and $\alpha+\lambda_0\xi_n$ falls into the same interval
as $E_{r_n+i}$ or $E_{r_n+j}$, can be made arbitrarily close to
one as well.

It is therefore enough to consider the intersection of the event
$Q_{ij}\geq \beta$, the event that
$E_{r_n+i}$ and $E_{r_n+j}$ fall into two different intervals
of the form $[\alpha+(m-1)\eta\xi_n,\alpha+m\eta\xi_n]$,
$m=1,\dots,\lambda_0/\eta$, and the event that both of them
are the only energies that fall into these intervals.
Denote the intersection of these events by
$A_{ij}(\beta)$.  Decomposing the event
$A_{ij}(\beta)$ according to the spin configurations
$\bss^{(r_n+i)}$ and $\bss^{(r_n+j)}$
corresponding to the
$i^{\text{th}}$ and $j^{\text{th}}$ energy above $\alpha$
and the particular intervals containing these energies, we then
rewrite the probability of $A_{ij}(\beta)$ as
\begin{equation}
\begin{aligned}
\PR(A_{ij}(\beta))=
\sum_{m_i<m_j}
\sumbeta_{\bss,\tilde\bss}
\PR\Bigl[
&A_{m_i}(\bss)\cap A_{m_j}(\tilde\bss)\cap\{Z_n^{(1)}=i-1\}
\cap\{Z_n^{(2)}=1\}
\\
&
\cap\{Z_n^{(3)}=j-i-1\}
\cap\{Z_n^{(4)}=1\}
\Bigr].
\end{aligned}
\end{equation}
Here the second sum runs over pairs of distinct configurations
$\bss,\tilde\bss$ with rescaled overlap larger than $\beta$,
the
first sum runs over  integers $m_i,m_j$ with
$0<m_i<m_j\leq \lambda_0/\eta$, the symbol
$A_m(\bss)$ denotes the event that the energy of the
configuration $\bss$ falls into the interval
$[\alpha+(m-1)\eta\xi_n,\alpha+m\eta\xi_n]$, and the
random variables $Z_n^{(\ell)}$
are equal to the number of points in the spectrum that
lie in the intervals $[a_n^{(\ell)},b_n^{(\ell)}]$ where
$a_n^{(1)}=\alpha$,
$b_n^{(1)}=a_n^{(2)}=\alpha+(m_1-1)\eta\xi_n$,
$b_n^{(2)}=a_n^{(3)}=\alpha+m_1\eta\xi_n$,
$b_n^{(3)}=a_n^{(4)}=\alpha+(m_2-1)\eta\xi_n$,
and
$b_n^{(4)}=\alpha+m_2\eta\xi_n$.
Defining
$I^{(\ell)}(\cdot)$ as before, let
\eq
\label{Z-overlap}
(Z_n)_2^{(\beta)}=
\sumbeta_{\bss,\tilde\bss}
I^{(2)}(\bss)I^{(4)}(\tilde\bss)
\en
be the number of distinct pairs of configurations
$\bss$, $\tilde\bss$ with rescaled overlap at least
$\beta$ such that the energy of $\bss$ falls into the interval
$[a_n^{(2)},b_n^{(2)}]$, and the energy of
$\tilde\bss$ falls into the interval
$[a_n^{(4)},b_n^{(4)}]$.  We then rewrite the probability
$\PR(A_{ij}(\beta))$ as
\begin{equation}
\label{overlap-rep2}
\begin{aligned}
\PR(A_{ij}(\beta))=
\sum_{m_i<m_j}
\EX\Bigl[
&(Z_n)_2^{(\beta)}
\I(Z_n^{(1)}=i-1)
\I(Z_n^{(2)}=1)
\\
&
\I(Z_n^{(3)}=j-i-1)
\I(Z_n^{(4)}=1)
\Bigr],
\end{aligned}
\end{equation}
where $\I(A)$ denotes the indicator function of the event
$A$.

Let $N_n(\beta)$ be the number of distinct pairs $\bss,\tilde\bss$
with rescaled overlap at least $\beta$.
Combining the methods of the last section with the standard
central limit theorem, we now
easily establish that
\eq
\label{overlap-con1}
\EX\Bigl[(Z_n)_2^{(\beta)}\Bigr]
=\eta^22^{-2n}N_n(\beta)\Big(1+o(1)\Big)
=\eta^2 R(\beta)\Big(1+o(1)\Big).
\en
In order to analyze the right hand side of
\eqref{overlap-rep2} we would like first to factor the
expectation on the right hand side, and then use
\eqref{overlap-con1} and Poisson
convergence of the random variables
$Z_n^{(\ell)}$
to analyze the resulting terms.  In the process,
we will have to analyze the factorial moments
\begin{equation}
\EX\Bigl[
(Z_n)_2^{(\beta)}
(Z_n^{(1)})_{k_1}
(Z_n^{(2)})_{k_2}
(Z_n^{(3)})_{k_3}
(Z_n^{(4)})_{k_4}
\Bigr].
\end{equation}
Unfortunately, the methods of the last section
cannot be directly applied to these factorial
moments
since the sum over configurations representing the above
expression
is not a sum over pairwise distinct configurations: comparing,
e.g., the sum over $\bss$ in \eqref{Z-overlap}
and the representation of
the random variable
$Z_n^{(2)}$ as a sum over configurations,
\begin{equation}
Z_n^{(2)} = \sum_{\bss'} I^{(2)}(\bss'),
\end{equation}
we see that both involve configurations
whose energy lies in the interval $[a_n^{(2)},b_n^{(2)}]$.
But this problem can be easily overcome by considering
the random variables $Z_n^{(2)}-1$
and $Z_n^{(4)}-1$ instead of $Z_n^{(2)}$
and $Z_n^{(4)}$.  We therefore consider the
expression
\eq
\begin{aligned}
\EX&\Bigl[
(Z_n)_2^{(\beta)}
(Z_n^{(1)})_{k_1}
(Z_n^{(2)}-1)_{k_2}
(Z_n^{(3)})_{k_3}
(Z_n^{(4)}-1)_{k_4}
\Bigr]
\\
&=
\sumbeta_{\bss,\tilde\bss}
\EX\Bigl[
I^{(2)}(\bss)I^{(4)}(\tilde\bss)
(Z_n^{(1)})_{k_1}
(Z_n^{(2)}-1)_{k_2}
(Z_n^{(3)})_{k_3}
(Z_n^{(4)}-1)_{k_4}
\Bigr].
\end{aligned}
\en
We claim that this expression can again be
be expressed as a double sum over distinct configurations,
allowing us to apply the methods of the last section.
Indeed, let us first consider the product
\begin{equation}
I^{(2)}(\bss)(Z_n^{(2)}-1)_{k_2}
=I^{(2)}(\bss)(Z_n^{(2)}-1)(Z_n^{(2)}-2)\cdots (Z_n^{(2)}-k_2).
\end{equation}
Proceeding as in the proof of \eqref{Zm}, we now rewrite this
product as a sum of configurations $\bss^{(1)},\dots,\bss^{(k_2)}$
which are mutually distinct and distinct from $\bss$.
In a similar way, the product $I^{(4)}(\tilde\bss)(Z_n^{(4)}-1)_{k_4}$
can be expressed as a sum over mutually distinct configurations which are
distinct from $\tilde\bss$.  Using these facts, we now proceed as
before to obtain the bound
\begin{equation}
\label{overlap-con2}
\begin{aligned}
\EX\Bigl[
(Z_n)_2^{(\beta)}&
(Z_n^{(1)})_{k_1}
(Z_n^{(2)}-1)_{k_2}
(Z_n^{(3)})_{k_3}
(Z_n^{(4)}-1)_{k_4}
\Bigr]
\\
&=
\eta^2\gamma_1^{k_1}\eta^{k_2}
\gamma_3^{k_3}\eta^{k_4}2^{-2n}N_n(\beta)\Bigl(1+o(1)\Bigr),
\end{aligned}
\end{equation}
where $\gamma_1=\eta(m_1-1)$ and $\gamma_3=\eta(m_2-m_1-1)$.

Consider the
four random variables
$Z_n^{(1)}$, $Z_n^{(2)}-1$,
$Z_n^{(3)}$ and $Z_n^{(4)}-1$, together with the probability
distribution $\mu$ defined by
\begin{equation}
\begin{aligned}
\mu\Bigl(
&Z_n^{(1)}=i_1,
Z_n^{(2)}-1=i_2,\,
Z_n^{(3)}=i_3,\,
Z_n^{(4)}-1=i_4\Bigr)
\\
&=
\frac{
\EX\Bigl[
(Z_n)_2^{(\beta)}
\I(Z_n^{(1)}=i_1)
\I(Z_n^{(2)}-1=i_2)
\I(Z_n^{(3)}=i_3)
\I(Z_n^{(4)}-1=i_4)
\Bigr]}
{\EX\Bigl[
(Z_n)_2^{(\beta)}
\Bigr]}.
\end{aligned}
\end{equation}
The bounds \eqref{overlap-con1} and \eqref{overlap-con2}
then establish that in the measure $\mu$, the four random
variables $Z_n^{(1)}$, $Z_n^{(2)}-1$,
$Z_n^{(3)}$ and $Z_n^{(4)}-1$ converge to
four independent Poisson random variables with rates
$\eta$, $\gamma_2$, $\eta$ and $\gamma_4$, respectively.
Using once more the bound \eqref{overlap-con1}, we conclude that
the expectation in the sum in \eqref{overlap-rep2}
can be approximated as
\begin{equation}
\label{overlap-rep4}
\begin{aligned}
\EX\Bigl[
&(Z_n)_2^{(\beta)}
\I(Z_n^{(1)}=i-1)
\I(Z_n^{(2)}-1=0)
\I(Z_n^{(3)}=j-i-1)
\I(Z_n^{(4)}-1=0)
\Bigr]
\\
&
=\eta^2 \frac{\gamma_1^{i-1}}{(i-1)!}
\frac{\gamma_3^{j-i-1}}{(j-i-1)!}
e^{-(2\eta+\gamma_1+\gamma_3)}
R(\beta)\Big(1+o(1)\Big).
\end{aligned}
\end{equation}
Inserted into \eqref{overlap-rep2} this gives
the bound
\begin{equation}
\label{overlap-rep3}
\PR(A_{ij}(\beta))=
K_\eta(\lambda_0)R(\beta)\Big(1+o(1)\Big),
\end{equation}
where
\begin{equation}
\begin{aligned}
K_\eta(\lambda_0)=
\eta^2\sum_{m_1<m_2} \frac{((m_1-1)\eta)^{i-1}}{(i-1)!}
\frac{((m_2-m_1-1)\eta)^{j-i-1}}{(j-i-1)!}
e^{-\eta m_2}
\end{aligned}
\end{equation}
is the Riemann-sum approximation to the integral
\begin{equation}
K(\lambda_0)=
\int_{0}^{\lambda_0}d\gamma_1
\frac{\gamma_1^{i-1}}{(i-1)!}e^{-\gamma_1}
\int_{0}^{\lambda_0}
d\gamma_3
\frac{\gamma_3^{j-i-1}}{(j-i-1)!}
e^{-\gamma_3}.
\end{equation}
As $\eta\to 0$, the Riemann sum
$K_\eta(\lambda_0)$ converges to the integral $K(\lambda_0)$,
and as $\lambda_0\to\infty$, the integral $K(\lambda_0)$
converges to $1$.  Choosing first
$\lambda_0$ large enough, then $\eta$ small enough, and then
$n$ large enough, the normal distribution function $R(\beta)$
is therefore an arbitrarily good approximation to
$\PR(A_{ij}(\beta)$, which in turn can be made arbitrary
close to $\PR(Q_{ij}\geq \beta)$,
again by first choosing $\lambda_0$ sufficiently large, then
$\eta$ sufficiently small, and then $n$ sufficiently large.
This establishes \eqref{PQ-lim} and hence the remaining statements
of Theorem~\ref{mainth}.

\section{Generalizations and open problems}

\subsection{Generalizations of the \npp}

The \npp\ has a natural generalization: Divide a set $\{X_1, X_2,
\dots, X_n\}$ of numbers into $q$~subsets such that the sums in all
$q$ subsets are as equal as possible. This is known as multi-way
partitioning or multiprocessor scheduling problem
\cite{bauke:mertens:npp}.  The latter name refers to the problem of
distributing $n$ tasks with running times $\{X_1,X_2,\dots,X_n\}$ on
$q$ processors of a parallel computer such that the overall running
time is minimized. Bovier and Kurkova \cite{bovier:kurkova:04}
considered the restricted multi-way partitioning
problem where the cardinality of each subset is fixed to $n/q$.
For this model they could prove the ``energy part'' of the local
REM hypothesis at $\alpha=0$, i.e., the convergence of the
properly scaled near optimal solutions to a Poisson point process.
The local REM (including the ``overlap part'') is conjectured to be valid
for all $\alpha \geq 0$ for the multi-way partitioning
problem in the unrestricted case (i.e., for $n/q$ not necessarily
fixed) \cite{rem1}.  This generalization is still open.

\subsection{Universality}

In \cite{rem2} it is conjectured that the local REM is a property of
discrete, disordered systems well beyond number partitioning and its
relatives.  Since this conjecture represents a fascinating open
problem for the rigorous community, we briefly review the heuristic
argument of \cite{rem2}: Consider a model with an energy function of
the form
\begin{equation}
  \label{universal1}
  E(\bss) = \sum_{i=1}^n \sigma_i X_i\,,
\end{equation}
where the $\bss$ is an $n$-dimensional vector with binary entries
$\sigma_i=\pm1$ or $\sigma_i\in\{0,1\}$ and the $X_i$ are real random
numbers from the unit interval.  In case of the \npp\ (or the $1$-d
Edwards-Anderson model), any vector $\bss$ is a feasible
configuration. If we add more restrictions, we could write the cost
function of many optimization problems in the form
(\eqref{universal1}).  For example, in the traveling salesman problem,
we would take $\sigma_i\in\{0,1\}$, where $\sigma_i=1$ means that the
distance $X_i$ is part of the tour, and the $\sigma_i$ would have to
fulfill the constraint to encode a valid itinerary. In
higher-dimensional spin glasses, the $\sigma_i=\pm 1$ encode satisfied
or unsatisfied edges, and are correlated due to loops in the graph.
In all cases we have an exponential number of valid configurations,
with an exponential number of energy values $E(\bss)$. Since the range
of energies scales only linearly with $n$, it should follow that
adjacent levels will be separated by exponentially small distances.
The \emph{precise} value of each gap will be determined by the
\emph{least significant bits} in the $X_i$'s, however. The dynamical
variables $\sigma_i$ can only control the $n$ most significant bits of
the energy.  \cite{rem2} argue that the residual entropy of the least
significant bits then gives rise to the Poisson nature of adjacent
energy levels and to the full local REM property. This very heuristic
argument has been supported by extensive numerical simulations in
various spin glass models (Edwards-Anderson model,
Sherrington-Kirkpatrick model, Potts glasses) and in optimization
problems (TSP, minimum spanning tree, shortest path) \cite{rem2}.

In this paper, we rigorously established the local REM conjecture for
a particular model, the \npp.  In a recent paper
\cite{bovier:kurkova:05}, submitted shortly after the present one,
Bovier and Kurkova showed that the local REM conjecture holds for many
types of spin glasses as well, in particular the Edwards-Anderson
model and the Sherrington-Kirkpatrick model.  Their approach is based
on a general theorem establishing Poisson convergence for an abstract
class of models, with \emph{conditions} that are very similar to the
\emph{statements} of our Lemmas~\ref{lem:Zkk-asymp} and
\ref{lem:comb-bds} in an abstract setting.

\subsection{Phase Transition}

According to the heuristic argument above, the bit-en\-tro\-py of the
disorder $X_i$ is the essential property that leads to the local REM:
if it is larger than the entropy of the configurations, the local REM
should apply. If it is lower than the configurational entropy, the
distances between adjacent energy levels are multiples of a fixed,
smallest distance. In this case, each energy level is populated by an
exponential number of configurations. An indicator for the transition
between the two regimes is the maximum overlap between two
configurations with adjacent energy levels. If the entropy of the
disorder is larger than the configurational entropy, this overlap
should be $0$ (the local REM).  If the entropy of the disorder is much
smaller than the configurational entropy, this overlap should be
$1-\Theta(n^{-1})$.  Numerical simulations in \cite{rem2} indicate
that there is a sharp transition at the point at which these entropies
are the same.  A canonical problem in which such a transition has been
rigorously investigated is the phase transition of the \npp\
\cite{borgs:chayes:pittel:01}.  \cite{rem2} propose that a transition
of this type may be as universal as the local REM.  A proof of the
universality of this transition poses yet another challenge for the
rigorous community.

\bigskip
{\em Acknowledgement:}
S.M.~was supported in part by the German Science Council (grant
\mbox{ME2044/1-1}), and C.N.~was supported by the Microsoft
Graduate Fellowship.  S.M.~would also like to thank Microsoft
Research for its hospitality.


\end{document}